# New perspectives for the fitness fatigue model: how to revisit questions about the science of sports training from the perspective of systems control theory

Jacky Montmain[1], Pierre Couturier[1], Gérard Dray[2]

[1] SyCoIA, IMT Mines Ales, France
jacky.montmain@mines-ales.fr, pierre.couturier@mines-ales.fr
[2] EuroMov Digital Health in Motion, Univ Montpellier, IMT Mines Ales, Ales, France
gerard.dray@mines-ales.fr

**Abstract**. The Fitness-Fatigue model (FFM) was initially designed to gain a better physiological understanding of the impact of training loads on sports performance. For almost 50 years, simulations have been compared with the observed response of sports performance to training loads. Understanding the relationship between training load and performance should have answered some fundamental questions for the physical trainer or sports coach: 1/ how to define the best training to achieve a performance without exhausting an athlete or how to achieve a performance in a limited time; 2/ how to assess the athlete's fatigue and fitness reliably to better understand his performance; 3/ how to prevent the risk of injury. But this was not the case. Studies dedicated to the FFM have been limited to improving somewhat on Banister's initial model without really taking the necessary step back to take advantage of the mathematical framework offered by the state representation, the implicit formalism underlying the FFM. The idea behind this research strategy is that having a valid and accurate model makes it easy to address previous questions through simulation: multiple training scenarios can be simulated until the ideal scenario for a given training question is identified. The main drawback to this approach is the combinatorial nature of the exercise. This paper is not discussing the relevance of the model, but how to use it. The state representation makes it possible to study the controllability, observability and diagnosability of a system (*i.e*, the athlete) and thus to formally answer the three previous practical questions. It can be considered that sports science studies have missed the richness of the FFM model. Artificial learning approaches are increasingly preferred to the FFM model because they are supposed to better capture observation. However, in the light of the state representation, the FFM model could still be extended naturally while remaining mathematically interpretable, offering mathematical tools to estimate the state of the athlete, and to define the most adequate training and prevent the risk of injury. This article does not aim to solve the question of optimal training in practice, for that it would need to be validated by specialists in physiology and sports science, it just proposes to take a fresh look at training issues and to demystify the mathematical formalism adopted by Banister.

## 1. Introduction

In this paper, the point is not to approve or deny Banister's model (Banister *et al.*, 1975; 1985; 1999; Calvert *et al.*, 1976), but to illustrate what should be done with the tools of process control theory if one is able to relate loads and performance by means of a state representation. We are not discussing the relevance of the model, but how to use it. By fitness-fatigue model FFM we mean the original Banister model and the variants that have developed from it over several decades (Busso *et al.*, 1990, 1992,1994, 2002; Clarke *et al.*, 2013; Piatrikova *et al.*, 2021). Many application studies (Gouba *et al.*, 2013) and very exhaustive reviews have recently been published (Hemingway, 2021; Hemingway *et al.*, 2021; Swinton *et al.*, 2021). Subsequent studies have too often focused on the recurrence equations of the FFM model alone, introducing pure delays or some interaction between fatigue and fitness to better approximate observed behaviour. The FFM could have yet answered some fundamental practical questions: 1/ how to define the best training to achieve a performance without exhausting an athlete or how to achieve a performance in a limited time; 2/ how to assess the athlete's fatigue and fitness reliably; 3/ how to prevent the risk of injury or any abnormal behaviors. Nevertheless, researchers seem to have been blinded by this race for validation and have locked themselves into the straitjacket of the equation of the physiological model (equation 1). Few of these studies have finally taken the necessary step back to place this 1975 model in the formalism of the state representation (Swinton *et al.*, 2021). This is undoubtedly a regrettable trend because by abandoning some of the physiological interpretation of the initial model, richer models could have been constructed, albeit at the expense of the physiological interpretation, but more precise and nevertheless mathematically interpretable. Indeed, in Banister's original model, the physiologically interpretable transfer function is the difference of two first orders which, with the time constants and gains envisaged in practice for the domain, gives a second order system with no phase minimum.

**Example 1:** *The original Banister's model. The variation in performance over time* $\Delta P(t) = P(t) - P^*$ *is the normalized difference (introduction of gains* $k_g, k_h$*) of the convolution products (*) of the load* $\omega(t)$ *with fitness function* $g(t)$ *and the load with fatigue function* $h(t)$ *:*

$$\Delta P(t) = P(t) - P^* = k_g.g(t) - k_h.h(t) \text{ with } \begin{cases} g(t) = \omega(t) * e^{-\frac{t}{\tau_g}} = \int_{u=0}^{t} e^{-\frac{(t-u)}{\tau_g}} \omega(u)du \\ h(t) = \omega(t) * e^{-\frac{t}{\tau_h}} = \int_{u=0}^{t} e^{-\frac{(t-u)}{\tau_h}} \omega(u)du \end{cases} \quad (1)$$

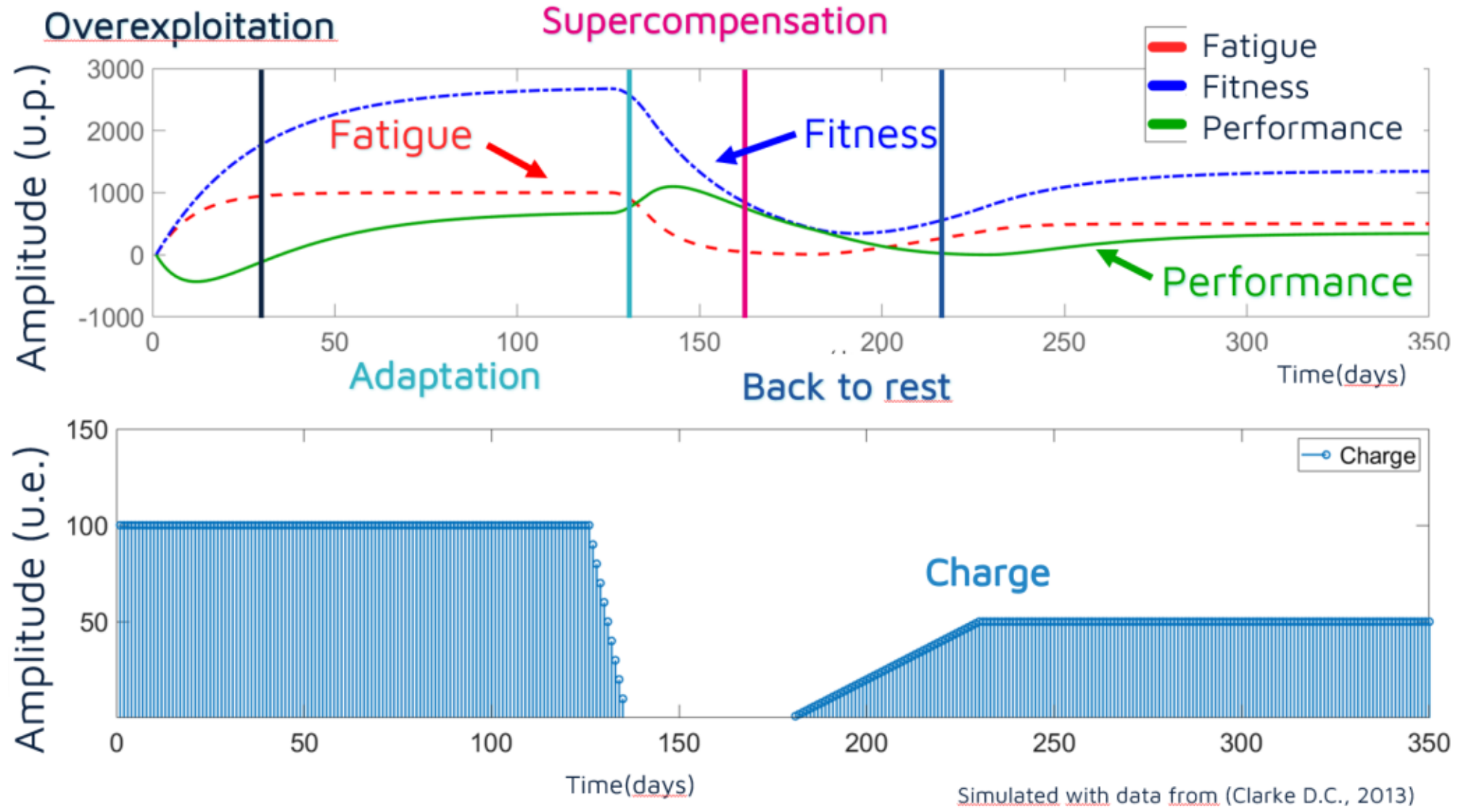

Fig. 1. Response to a train of charge doses

Throughout this study, this original FFM will be used as an illustration, but as mentionned above, it can be replaced by any other linear dynamic model. In addition, it is also possible to move away from the SISO (single input - single output, here charge and performance) model towards MIMO (multiple input - multiple output) model in order to no longer have to reduce the load and performance to a unique signal each (several types of training loads and various performance measures could be envisaged without further difficulty). Thus, the state representation could have opened up new horizons in terms of controllability and diagnosability for coaches and physical trainers. This study tries to give a new scope to the historical FFM.

The objective of this study is to show the functionalities offered by the algebraic formalism of the FFM model, which can be enriched at will by the state representation to better match the observation. The FFM model was originally designed to capture physiological behaviour (Hellard *et al*., 2006). The interpretability of the model took precedence over accuracy. Most of the time the FFM model is used in simulation to show that the model tracks well the performance associated with a given training. Trying to determine the training that would achieve a performance over a given time interval or minimize the fatigue of an athlete or the energy of the load signal requires testing multiple inputs until the one that satisfies the coach's or trainer's objective is found. Seen in this light, the problem of identifying the best training seems to be an insoluble combinatorial search problem. However, Banister's model, which is based on the mathematical concept of state representation, avoids this pitfall and offers many other possibilities than simple simulation! This is what we propose to show in this paper.

Today, sports researchers prefer models derived from machine learning to better match observations of sports performance (Imbach *et al*., 2022). However, these models are difficult to interpret and therefore do not lend themselves easily to the definition of adaptive training that is sensitive to the context or to the athlete's state of fitness. Moreover, machine learning models require large amounts of data before they can become effective simulation tools. This study shows that the state representation used by specialists in process control would also make it possible to generalise and render more faithful the FFM simulations of performance while offering effective tools for adapting training to the state of the athlete, diagnosing risks of injury or correcting the uncertainties of measurements.

Extensions of FFM models via state representation would allow:

- to propose two different visions of time in modelling, discrete time and continuous time, to show how one switches from a recurrence equation to a transfer function in $z$ (z-transform) or in $p$ (Laplace transform). This study should allow to demystify and understand the origin of the Banister model and to take a step back to imagine more complex models if necessary according to the same modelling principles
  → generalization of the FFM to any linear dynamic model through the state representation in continuous or discrete time;
- to show how to calculate the parameters of a linear dynamic model with a simple least squares algorithm and few training points. The identification of the model is thus an infinitely less demanding process than the algorithms of machine learning
  → frugal identification of linear dynamic models based on a simple least squares algorithm;
- to show how to construct an observer of the state of a system when one has a linear dynamic model and measurements. Fatigue and fitness are not measured quantities, the observer makes it possible to estimate them in real time thanks to the dynamic model and consequently to compute the best training load in order the performance to progress according to the trainer's objectives
  → reliable estimation of the athlete's fatigue and fitness without assumption on the initial values;
- to show how the training load can be adapted to the athlete's physical condition to achieve a given level of performance by optimising a given energy criterion or by taking into account the athlete's condition, *i.e.*, fatigue and fitness. This study should show how training can be defined according to criteria such as minimising the training load or the athlete's fatigue, how reducing the response time to reach a stable performance faster or according to specifications on the dynamics of the evolution of the performance
→ state feedback control, optimal control and pole placement control to define the best training according to the athlete's form and fatigue;
- to show how to generate analytical redundancy relationships involving the available measures to detect anomalies in the athlete's behaviour at an early stage and thus prevent unwanted events such as the risk of injury
→ analytical redundancy relationships are virtual sensors dedicated to fault detection and isolation to prevent injuries;
- to show how to incorporate model or measurement uncertainties into the model
→ a Kalman filter allows to correct the measurement uncertainties to reliably follow the athlete's behavior.

The highlights of this study could be summarized as follows:

- to demystify the FFM by introducing a mathematical formalisation based on state space representation and linear process control;
- to analyse the question of finding the optimal training as a control problem;
- more generally, to open up new perspectives on the use of the FFM by formalising sports science issues in terms of mathematical problems of control theory.

The following table summarises the contribution of this study and announces the formalisation that will be used later in this study to transcribe problems from training science into mathematical problems of control theory:

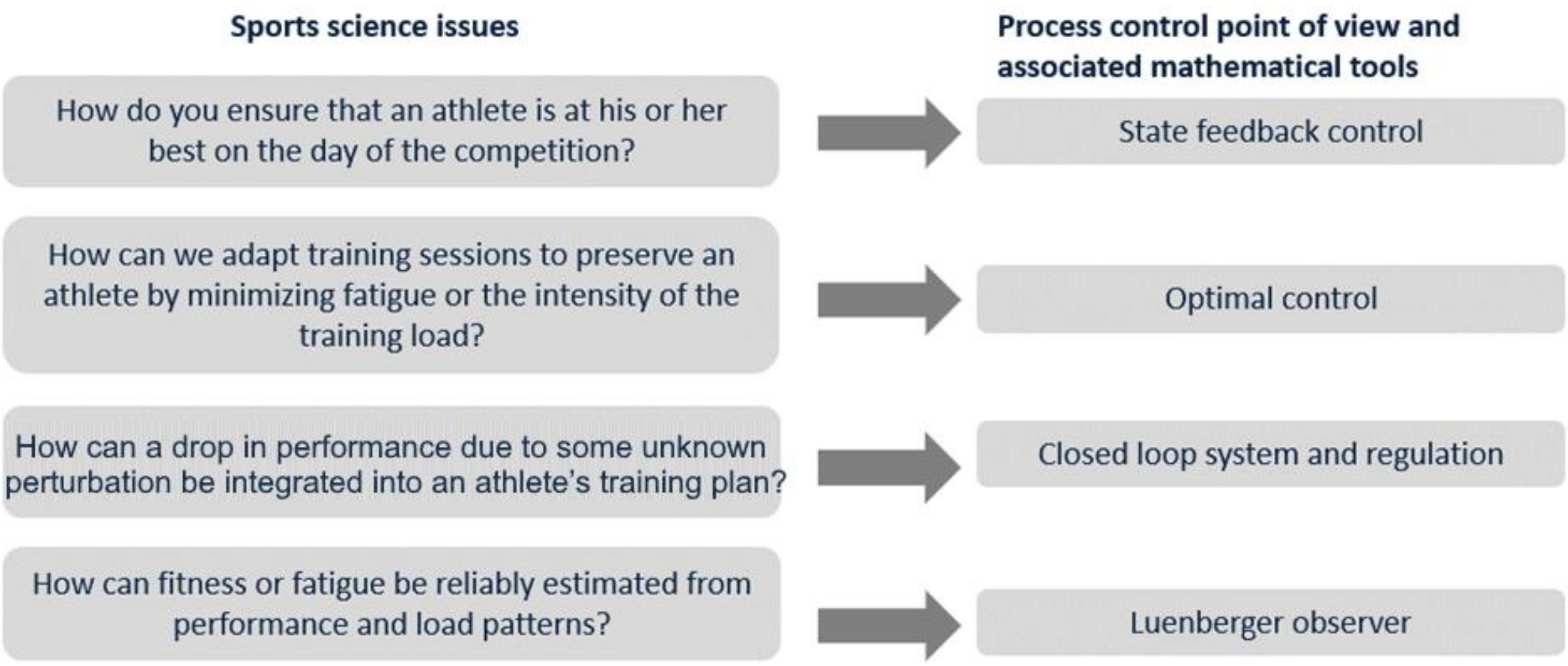

Before sport science researchers abandoned the FFM model, our study aims on the one hand to explain that while remaining in the same formalism of the state representation the precision of the model could clearly be improved by choosing more complex algebraic structures but which remain interpretable, and on the other hand to show all the richness of the state representation in terms of functionalities which could be directly usable by a coach or a physical trainer. So before burying the FFM model for pure machine learning techniques, this study proposes to take a good look at all the richness of this type of model. It goes without saying that our study would require physiological validation before it could be considered in practice, but that is not our purpose here, we simply want to encourage sports trainers to ask themselves training questions from a control theory perspective.

The outline of this paper is as follows. Section 2 reviews the mathematical foundations of Banister's model, makes explicit the state representation in continuous and discretised time. Section 3 explains the least squares algorithm for the identification of linear dynamic models like Banister's. Section 4 is dedicated to the control based on a state representation, it opens new perspectives for the determination of the most appropriate training for an athlete's state. Section 5 presents the concept of an observer that allows the FFM model to correct its predictions from observations or to no longer make assumptions about the initial values of fatigue and fitness. Section 6 introduces the concept of diagnosability of dynamical systems: this concept could help to prevent injury risks. Section 7 describes how the Kalman filter can be combined with a state-space representation of a FFM to obtain better estimates of fitness and fatigue with uncertain incoming data. A conclusion looks at the new perspectives that the FFM could offer in terms of functionalities to coaches and trainers.

## 2. State representation

There are several advantages to using the state representation formalism and the related transfer functions as a model for the relationships between the training loads and the fitness or fatigue variables exploited in FFM related papers.

Firstly, Banister's original model corresponds to the difference of two simple first order transfer functions (equation 1), but it could be extended to more complex transfer functions to better match performance observations. This would allow much more sophisticated dynamic relationships between loads and state variables to be modelled. The idea of this section is to provide an understanding of the origin of the equations proposed by Banister so that they can be easily extended to more sophisticated models. The only drawback of this state representation-based extension would be the loss of direct physiological interpretation, but the model identification phase would not be more complicated. Furthermore, load and performance are two real signals that could be extended to input and output vectors (several types of training loads could be defined and various performance measures could be collected) if one wished to have a multidimensional view of both load and performance. This extension to the multidimensional case of load and performance presents no difficulty in the state representation framework.

Secondly, transfer functions are the basic tools of control theory. Optimal control theory is a branch of mathematical optimisation that consists of finding a control for a dynamic system over a period of time such that an objective function is optimised. In classical linear quadratic (LQ) optimal control problems, the resulting control law (*i.e*. the training loads in our case) can be provided analytically from the algebraic structure of the dynamic system and the expected output. The optimal control law is a time-varying linear function of the state variables. The control theory framework thus offers the possibility to design the optimal training loads to be planned in order to reach a given performance set point while minimising some energy criteria that can be considered as a weighted function on state variables (*e.g*. fitness, fatigue, etc.) and inputs (*e.g*. training load doses) in LQ problems. This is a trade-off between energy cost and response time of the closed-loop system, and depends on the vision of the decision maker (*e.g*. the fitness trainer's strategy). However, although the first FFM models appeared almost 50 years ago, their use to design optimal training load planning is systematically considered through simulations to our knowledge, whereas their main advantage lies in their algebraic structure for control purposes which avoids the combinatory of simulations!

Finally, let us add that the untapped algebraic structure of FFM models would also provide state observers of fitness and fatigue. A state observer is a system that provides an estimate of the internal state of a given real system, based on measurements of the input and output of the real system. In our field of study, it could be used to accurately estimate the state variables of athletes (fitness and fatigue) or to readjust the model through performance observations. One could have a systematic estimation of the state of fatigue and compare it for example to the subjective Rating of Perceived Exertion (RPE - level or index of perceived effort). The algebraic structure of a FFM model also allows the generation of analytical redundancy relationships between measured quantities which are indicators of anomalies in the athlete's behaviour, which could be used to prevent a risk of injury or counter-performance. Lastly, the modelling of measurement or model uncertainties is done naturally by adding noise to the state and observation equations.

This section therefore introduces the state representation, which constitutes the formal framework of Banister's model; from this theoretical framework, the different control or diagnosis functionalities useful to the physical trainer or the sports coach will be developed in the following sections. The following sections address these issues with the state representation formalism, which we believe should have been introduced in the early stages of the FFM model to avoid confusion and limit the almost hieratic proliferation of Banister model extensions.

## 2.1. Continuous time modelling

A dynamic model of a continuous system can be written as:

$$\frac{dX(t)}{dt} = f\left(X(t), U(t)\right) \tag{2}$$

$$Y(t) = v\left(X(t), U(t)\right) \tag{3}$$

where $U(t)$ is the input to the system and $Y(t)$ the output, *i.e.* the quantities of interest for the study and $X(t)$ the vector of state variables. The output values are measured while the input values may be adjusted to vary the output. State variables are generally not measured for reasons of infeasibility or cost. A single-variable system has only one input signal and one output signal.

If the dynamic model is linear and time invariant equations (2) and (3) become:

$$\frac{dX}{dt} = AX(t) + BU(t) \tag{4}$$

$$Y(t) = CX(t) + DU(t) \tag{5}$$

$U(t)$, $Y(t)$ and $X(t)$ are vectors in the general case.

Equation (4) is called the state equation and is the differential equation that defines the dynamic behavior of the system. Equation (5) is called the observation equation. $C$ is the observation matrix. In general, it is difficult or expensive to measure all the state variables. The use of the model to make predictions at a given time horizon is therefore based on the state equation.

If the system is single-variable, linear and time invariant (which is the case with the Banister model), then the input output differential equation can be written as:

$$\frac{d^n y(t)}{dt^n} + a_{n-1}\frac{d^{n-1}y(t)}{dt^{n-1}} + a_{n-2}\frac{d^{n-2}y(t)}{dt^{n-2}} + .. + a_0 y(t) = b_0 u(t) + b_1\frac{du(t)}{dt} + .. + b_m\frac{d^m u(t)}{dt^m} \tag{6}$$

with $m \leq n$

A possible state representation for the system described by (6) is:

$$\frac{dX(t)}{dt}=\frac{d}{dt}\begin{bmatrix} X(t) \\ dX/dt \\ d^2X/dt^2 \\ .. \\ dX^{n-1}/dt^{n-1} \end{bmatrix}=\begin{bmatrix} 0 & 1 & 0 & .. & 0 \\ 0 & 0 & 1 & .. & 0 \\ 0 & 0 & 0 & .. & 0 \\ .. & .. & .. & .. & .. \\ -a_0 & -a_1 & -a_2 & .. & -a_{n-1} \end{bmatrix}\begin{bmatrix} X(t) \\ dX/dt \\ d^2X/dt^2 \\ .. \\ dX^{n-1}/dt^{n-1} \end{bmatrix}+\begin{bmatrix} 0 \\ 0 \\ 0 \\ .. \\ 1 \end{bmatrix}U(t) \tag{7}$$

$$Y(t)=\begin{bmatrix} b_0 & b_1 & .. & b_m & 0 & ... & 0 \end{bmatrix}X(t)$$

The Laplace transform can be introduced to compute the ratio $Y(p)/U(p)$ *that is defined for single input-single output system as the transfer function* $H(p)$ *between the output and the input of the system with null initial conditions. Equation (6) becomes:*

$$p^n y(p)+a_{n-1}p^{n-1}y(p)+a_{n-2}p^{n-2}y(p)+..+a_0y(t)=b_0u(t)+b_1pu(p)+..+b_mp^mu(p)$$

and: $$H(p)=\frac{Y(p)}{U(p)}=\frac{\left(b_0+b_1p+..+b_mp^m\right)}{\left(a_0+..+a_{n-2}p^{n-2}+a_{n-1}p^{n-1}+p^n\right)} \tag{8}$$

**Example 2:** *Let's go back to the Banister model with the state representation. The original Banister model will be used in each section to illustrate the concept introduced by the section.*

*The basic equations are: assuming* $k_h > k_g > 0$ *and* $\tau_g > \tau_h > 0$

$$\begin{cases} \dfrac{dg(t)}{dt}=\omega(t)-\dfrac{1}{\tau_g}.g(t) \\ \dfrac{dh(t)}{dt}=\omega(t)-\dfrac{1}{\tau_h}.h(t) \end{cases} \Leftrightarrow \begin{pmatrix} \dot{g}(t) \\ \dot{h}(t) \end{pmatrix}=\underbrace{\begin{pmatrix} -\dfrac{1}{\tau_g} & 0 \\ 0 & -\dfrac{1}{\tau_h} \end{pmatrix}}_{A}\begin{pmatrix} g(t) \\ h(t) \end{pmatrix}+\underbrace{\begin{pmatrix} 1 \\ 1 \end{pmatrix}}_{B}\omega(t)=A\begin{pmatrix} g(t) \\ h(t) \end{pmatrix}+B\omega(t) \tag{9}$$

$g(t)$ and $h(t)$ are the state variables of the system.

$$P=k_g.g(t)-k_h.h(t)=\begin{pmatrix} k_g & -k_h \end{pmatrix}\begin{pmatrix} g(t) \\ h(t) \end{pmatrix}=Cx(t) \text{ with } C=\begin{pmatrix} k_g & -k_h \end{pmatrix} \tag{10}$$

*In Laplace transform, with null initial conditions (i.e. for variations around an equilibrium point), we can write:*

$$\begin{cases} \dfrac{dg(t)}{dt}=\omega(t)-\dfrac{1}{\tau_g}.g(t) \\ \dfrac{dh(t)}{dt}=\omega(t)-\dfrac{1}{\tau_h}.h(t) \end{cases} \Leftrightarrow \begin{cases} pg(p)=\omega(p)-\dfrac{1}{\tau_g}.g(p) \\ ph(p)=\omega(p)-\dfrac{1}{\tau_h}.h(p) \end{cases} \Leftrightarrow \begin{cases} g(p)=\dfrac{1}{\left(p+\dfrac{1}{\tau_g}\right)}\omega(p) \\ h(p)=\dfrac{1}{\left(p+\dfrac{1}{\tau_h}\right)}\omega(p) \end{cases} \tag{11}$$

*We have the state* $\begin{pmatrix} g(t) \\ h(t) \end{pmatrix}$ *as a function of the input and if we multiply the state by* $C=\begin{pmatrix} k_g & -k_h \end{pmatrix}$*, we obtain the output as a function of the input using the Laplace transform.*

*And as a result:*

$$\Delta P(p)=\left(\frac{k_g}{\left(p+\frac{1}{\tau_g}\right)}-\frac{k_h}{\left(p+\frac{1}{\tau_h}\right)}\right)\omega(p) \tag{12}$$

$$\Rightarrow \Delta P(t)=P(t)-P^*=\omega(t)*\left(k_g e^{-\frac{t}{\tau_g}}-k_h e^{-\frac{t}{\tau_h}}\right)=\omega(t)*e^{-\frac{t}{\tau_g}}-\omega(t)*e^{-\frac{t}{\tau_h}}=g(t)-h(t)$$ [1]

*This is the equation of the original Banister's model (equation 1).*

*These equations allow prediction or simulation to calculate the state variables and then, if necessary, the outputs of the system with the observation equation.*

*When* $\omega(t)$ *is an impulse (what Banister calls doses seen as diracs),* $\omega(t)=\delta(t), i.e\ \delta(0)=1$ $if\ t=0,\ 0\ else$*, then* $\omega(p)=1$ and *we check the boundary conditions at* $t=0$ *and* $t\to\infty$*:*

$$\lim_{p\to\infty} p\Delta P(p)=\lim_{t\to 0}\Delta P(t)=(k_g-k_h) \text{ and } \lim_{p\to 0} p\Delta P(p)=\lim_{t\to\infty}\Delta P(t)=0 \tag{13}$$

## 2.2. Discrete time modelling

This view of continuous time allows us to find the equivalent of Banister's convolution products in Laplace very easily, but it is sometimes not easy to implement numerically, so we prefer to discretise time, *i.e.* time only takes discrete values $t=kT$ or $t=k$ when the sampling period $T=1$ (which will be the case in the simulations of this study $(T=1\ day)$ as assumed in Banister's model, without losing generality).

### 2.2.1. The input is seen as a train of pulses

This is the basic assumption of Banister's model: the training load is a train of training pulses or "doses".

**Example 3:** *The discretization of the above convolution product can be done directly. If the sampling period T is 1,* $g(t)=\omega(t)*e^{-\frac{t}{\tau_g}}=\int_{u=0}^{t} e^{-\frac{(t-u)}{\tau_g}}\omega(u)du$ *is written as:*

---

[1] * is the convolution product.

$$g(n)=\sum_{i=0}^{n}\omega(i).e^{-\frac{(n-i)}{\tau_g}}=\sum_{i=0}^{n}\omega(i).e^{-\frac{(n-1-i+1)}{\tau_g}}=e^{-\frac{1}{\tau_g}}\left(\sum_{i=0}^{n-1}\omega(i).e^{-\frac{(n-1-i)}{\tau_g}}\right)+\omega(n)=e^{-\frac{1}{\tau_g}}.g(n-1)+\omega(n)$$

(14)[2]

*The formulation of the Banister model in the form of a recurrence equation can be useful in simulation or prediction in particular.*

For the record, the discrete time representation is less abstract because if the input is a train of pulses, one way to interpret the convolution product is to sum the impulse responses to each of the pulses. The responses are all the same within one multiplier and one delay.

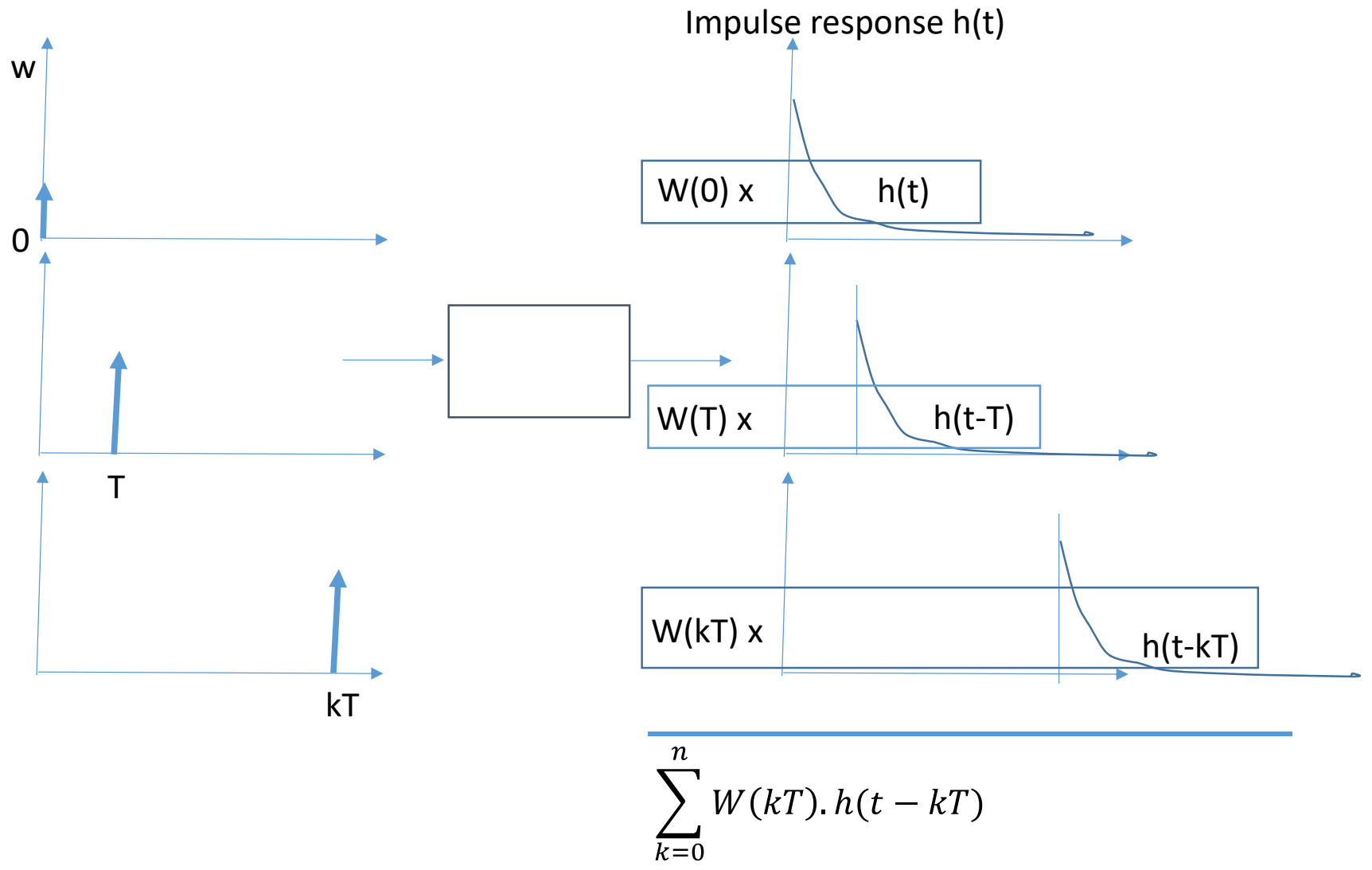

Fig. 2. Interpretation of the response to a pulse train by a linear system

Thus, $y(t)=\sum_i \omega(iT).h(t-iT)$ and if $T=1$: $y(n)=\sum_{i=0}^{n}\omega(i).h(n-i)$. This illustration allows us to interpret the concept of the convolution product used in the FFM model and to see that this principle is transposable to any transfer function.

### 2.2.2. With the z-transform

When working in discrete time, another way of proceeding is to use the z-transform and not the Laplace transform. The z-transform of a unit pulse is 1 and we have the table of transforms which gives us:

| Temporal signal | Laplace transform | z-transform |
|---|---|---|
| $e^{-at}$ | $\frac{1}{p+a}$ | $\frac{z}{z-e^{-aT}}$ |

[2] This equation is the mathematical discretization of the convolution product as conventionally done in signal processing. It is not quite the recurrence equation often found in the sports literature. It is, however, the theoretical calculation associated with this model and we re-demonstrate it below using the z-transform. We will see in the next section that by modelling the load no longer as a train of pulses, but as a constant piecewise signal, we find a more interpretable temporal causality.

**Example 4:** *Thus for the fitness function of the Banister model for example* $g(p)=F(p).\omega(p)$ $\Leftrightarrow g(z)=f(z).\omega(z)$ *, we calculate:*

$$TZ\left(g(p)\right)=TZ\left(\frac{1}{\left(p+\frac{1}{\tau_g}\right)}\right)\omega\left(z\right)=\frac{z}{z-e^{-\frac{1}{\tau_g}}}\omega\left(z\right)$$

*This gives the recurrence equation in discrete time:*

$$g\left(n+1\right)-e^{-\frac{1}{\tau_g}}g(n)=\omega\left(n+1\right) \text{ or } g\left(n\right)=e^{-\frac{1}{\tau_g}}g(n-1)+\omega\left(n\right)$$

*The recurrence equation (14) obtained by the direct discretization of continuous time in the example 3 can be easily found with z-transform.*

## 2.3. The input seen as a continuous signal

Another load model hypothesis would be to see the input not as a pulse train, but as an input variable with a constant value over a period between two drives. The load is then more realistically modelled as a sequence of slots (a piecewise continuous signal) and no longer as a pulse train. It is not very realistic to imagine that an athlete could take a dirac as a load, as proposed in the original Banister model. In this section, we propose some reminders to transform the input signal into a piecewise continuous signal.

For the record, let's consider an elementary slot of duration 1 (sampling period T=1) and its z-transform:

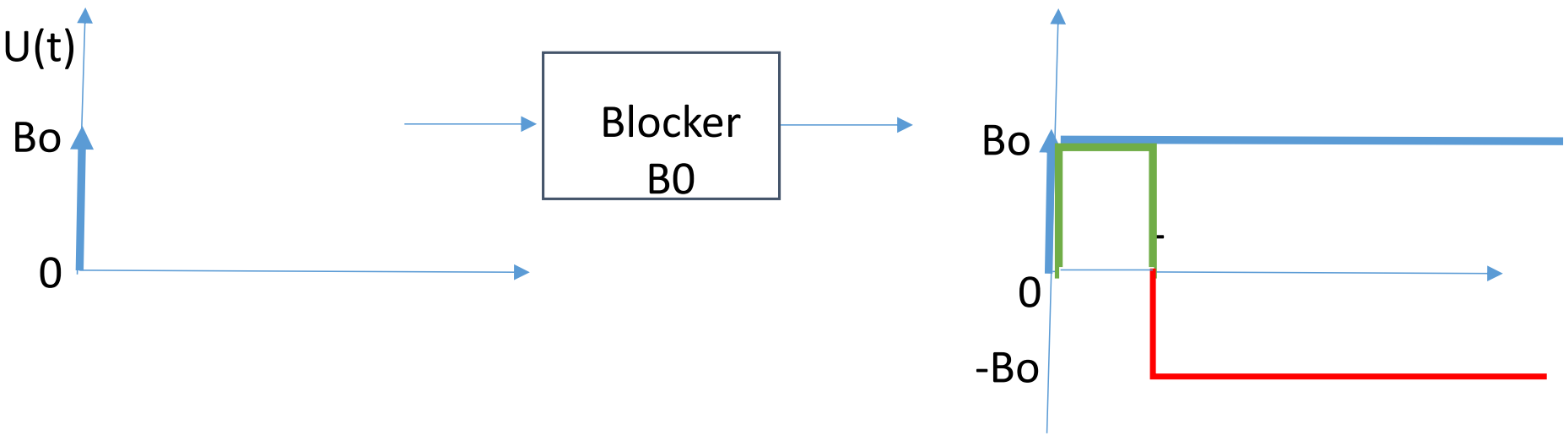

Fig. 3: From impulse to slots

The slot is seen as the sum of two steps (the first one starts at 0 (blue) and the delayed one (red) is of negative sign). Hence, the Laplace transform is:

$$TL\left(B_0\left(t\right)\right)=\frac{1}{p}-\frac{e^{-Tp}}{p}=\frac{1}{p}\left(1-e^{-Tp}\right)$$

Now, let's imagine that the load is a sequence of slots, *i.e.* we block the values of the pulse train over a sample period:

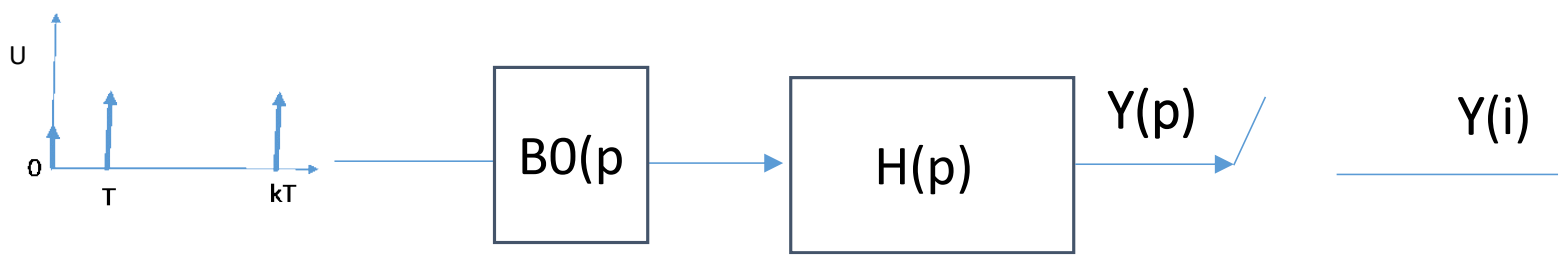

We then have:

$$G(z)=TZ\left(B_0(p)H(p)\right)=TZ\left(\left(1-e^{-Tp}\right)\frac{H(p)}{p}\right)=TZ\left(\frac{H(p)}{p}\right)-TZ\left(e^{-Tp}\frac{H(p)}{p}\right)$$
$$=TZ\left(\frac{H(p)}{p}\right)-z^{-1}TZ\left(\frac{H(p)}{p}\right)=\left(1-z^{-1}\right)TZ\left(\frac{H(p)}{p}\right) \quad (15)$$

Finally, equivalently to (8), $G(z)$ can be written as the ratio of two z-polynomials:

$$G(z)=\frac{N(z)}{D(z)} \quad \text{(z-transfer function)} \quad (16)$$

**Example 5:** *Thus, for the Banister model we have:*

$\Delta P(p)=\underbrace{\left(\frac{k_g}{\left(p+\frac{1}{\tau_g}\right)}-\frac{k_h}{\left(p+\frac{1}{\tau_h}\right)}\right)}_{H(p)}\omega(p)$ *and the table of transforms gives :*

| Temporal signal | Laplace transform | z-transform |
|---|---|---|
| $\left(1-e^{-at}\right).u(t)$ | $\frac{a}{p(p+a)}$ | $\frac{\left(1-e^{-aT}\right)z}{(z-1)\left(z-e^{-aT}\right)}$ |

$$TZ\left(B_0(p)H(p)\right)=TZ\left(B_0(p)\left(\frac{k_g}{\left(p+\frac{1}{\tau_g}\right)}-\frac{k_h}{\left(p+\frac{1}{\tau_h}\right)}\right)\right)=\left(1-z^{-1}\right)TZ\left(\frac{k_g}{p\left(p+\frac{1}{\tau_g}\right)}-\frac{k_h}{p\left(p+\frac{1}{\tau_h}\right)}\right)$$

$$\frac{\Delta P(z)}{\omega(z)}=\left(1-z^{-1}\right)TZ\left(H(p)\right)=\tau_g k_g\frac{\left(1-e^{-\frac{T}{\tau_g}}\right)}{\left(z-e^{-\frac{T}{\tau_g}}\right)}-\tau_h k_h\frac{\left(1-e^{-\frac{T}{\tau_h}}\right)}{\left(z-e^{-\frac{T}{\tau_h}}\right)} \quad (17)$$

$$\Rightarrow g(n)=e^{-\frac{T}{\tau_g}}g(n-1)+\tau_g\left(1-e^{-\frac{T}{\tau_g}}\right)\omega(n-1),\ h(n)=e^{-\frac{T}{\tau_h}}h(n-1)+\tau_h\left(1-e^{-\frac{T}{\tau_h}}\right)\omega(n-1)$$

*Note that these recurrence equations are obviously different from those of the Banister model established earlier. The z-transform is the most convenient tool to establish these recurrence equations which give the dynamics of a system.*

Another way of proceeding for discretisation with a continuous input is to discretise the differential equation directly over a sample period:

$$\frac{dX(t)}{dt} = AX(t) + BU(t)$$

The solution of the homogeneous differential equation without second member $BU(t)$ is: $X_H = Ce^{A(t-t_0)}$

To calculate the global solution, the method of variation of the constant is used:

$$C' e^{A(t-t_0)} = BU(t) \Rightarrow C' = e^{-A(t-t_0)} BU(t) \Rightarrow C = \int_{t_0}^{t} e^{-A(\tau-t_0)} BU(\tau) d\tau + c$$

$$X(t) = \underbrace{e^{A(t-t_0)} X(t_0)}_{c=X(t_0)} + e^{A(t-t_0)} \int_{t_0}^{t} e^{-A(\tau-t_0)} BU(\tau) d\tau = e^{A(t-t_0)} X(t_0) + \int_{t_0}^{t} e^{A(t-\tau)} BU(\tau) d\tau$$

If we discretise between two consecutive instants between which the input has remained constant (blocker principle) we obtain a recurrence equation which is none other than the discrete equation of state:

$$X\left((k+1)T\right) = e^{AT} X(kT) + \left(\int_{kT}^{(k+1)T} e^{A((k+1)T-\tau)} B d\tau\right) U_k$$
$$= e^{AT} X(kT) - \left(\int_{T}^{0} e^{A\theta} B d\theta\right) U_k = \underbrace{e^{AT}}_{\bar{A}} X(kT) + \underbrace{\left(\int_{0}^{T} e^{A\theta} B d\theta\right)}_{\bar{B}} U_k$$

Or:

The continuous model :

$$\underbrace{\begin{cases} \dot{X} = AX + BU \\ Y = CX + DU \end{cases}}_{State\ representation} \quad \underset{null\ initial\ conditions}{\Leftrightarrow} \quad \underbrace{\begin{cases} (pI-A)X(p) = BU(p) \\ Y(p) = \left(C(pI-A)^{-1}B + D\right)U(p) \end{cases}}_{transfer\ function}$$

Becomes discreet:

$$\underbrace{\begin{cases} X_{k+1} = \bar{A}X_k + \bar{B}U_k \\ Y_k = \bar{C}X_k + \bar{D}U_k \end{cases}}_{State\ representation} \quad \underset{null\ initial\ conditions}{\Leftrightarrow} \quad \underbrace{\begin{cases} (zI-\bar{A})X(z) = \bar{B}U(z) \\ Y(z) = \left(\bar{C}(zI-\bar{A})^{-1}\bar{B} + \bar{D}\right)U(z) \end{cases}}_{Transfer\ function}$$

**Exemple 6:** *This formula is applied to the Banister fitness function (or fatigue function):*

$$g\left((k+1)T\right) = e^{-\frac{1}{\tau_g}T} g(kT) - \tau_g \left[e^{-\frac{1}{\tau_g}\theta}\right]_0^T .1.\omega(kT) = e^{-\frac{1}{\tau_g}T} g(kT) - \tau_g \left[e^{-\frac{T}{\tau_g}} - 1\right] \omega(kT)$$
$$= e^{-\frac{1}{\tau_g}T} g(kT) + \tau_g \left[1 - e^{-\frac{T}{\tau_g}}\right] \omega(kT)$$

*or for* $T = 1$*:*

$$g(n) = e^{-\frac{1}{\tau_g}T} g(n-1) + \tau_g \left[ 1 - e^{-\frac{T}{\tau_g}} \right] \omega(n-1), \; h(n) = e^{-\frac{1}{\tau_h}T} g(n-1) + \tau_h \left[ 1 - e^{-\frac{T}{\tau_h}} \right] \omega(n-1)$$

*The same formulae are found as in the z-transform in the example 5. The z-transform allows us to obtain these relations more directly without doing integral calculations. These recurrence equations are used for prediction or simulation to calculate the state variables, and then if necessary the outputs of the system.*

**Example 7:** *The parameters we have chosen for this fitness-fatigue model simulations are commonly found in the literature:*

| $\tau_h$ | $\tau_g$ | $k_g$ | $k_h$ |
|---|---|---|---|
| 10 days | 30 days | 0.030 | 0.035 |

*The quantitative values of these parameters are not essential to this study, we simply wanted to take orders of magnitude that make sense for purely illustrative purposes of what can be achieved in control on this type of system. The following simulations were carried out in Matlab. The "athlete" system in open loop is then described by the following diagram:*

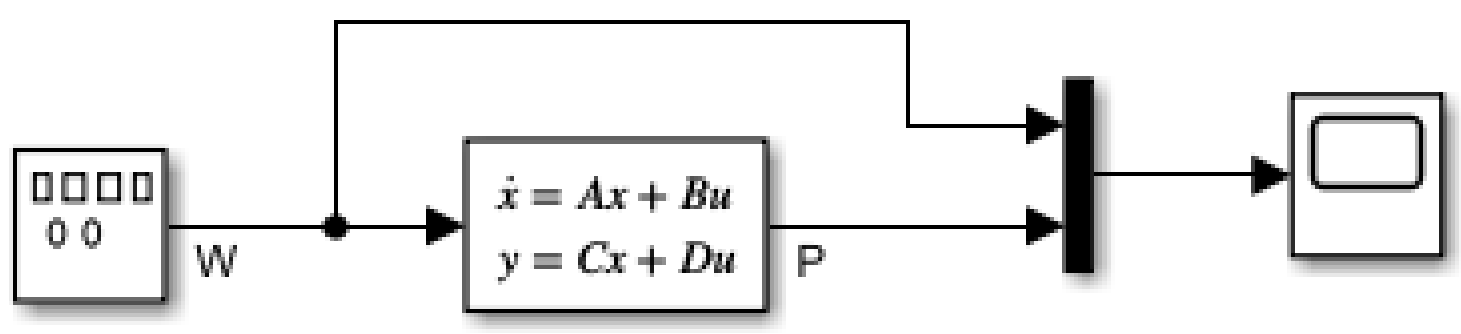

*Fig. 4:* *The 'athlete' system in open loop in Matlab/Simulink environment*

*The initial conditions of the system are usually set at* $g(0) = h(0) = 0$ *if the athlete has not exercised for a significant period of time, otherwise these initial values may have to be estimated a priori (see the section with an observer). Similarly, it is generally assumed that* $P(0) = 0$.*Null initial conditions refer also to variation studies around equilibrium points.*

*Under these conditions, the response to a pulse from the load* $w(t)$ *at* $t = 0$ *is as follows:*

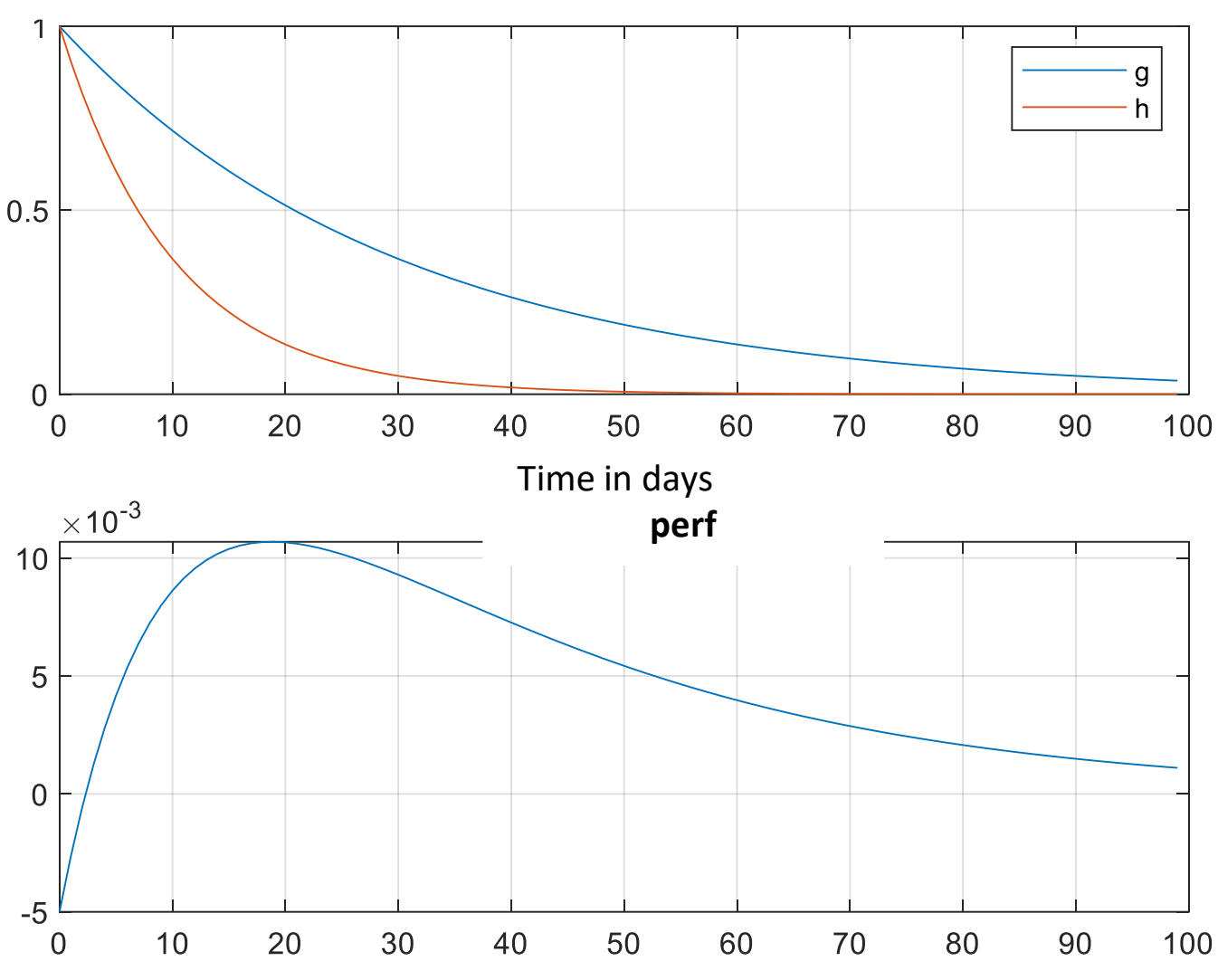

*Fig. 5: 100-day impulse response*

*It can be noted that after 100 days, the steady state is still not reached.*

*If now the input $w(t)$ is a pulse train of a 20-day period, i.e. there is a training session every 20 days:*

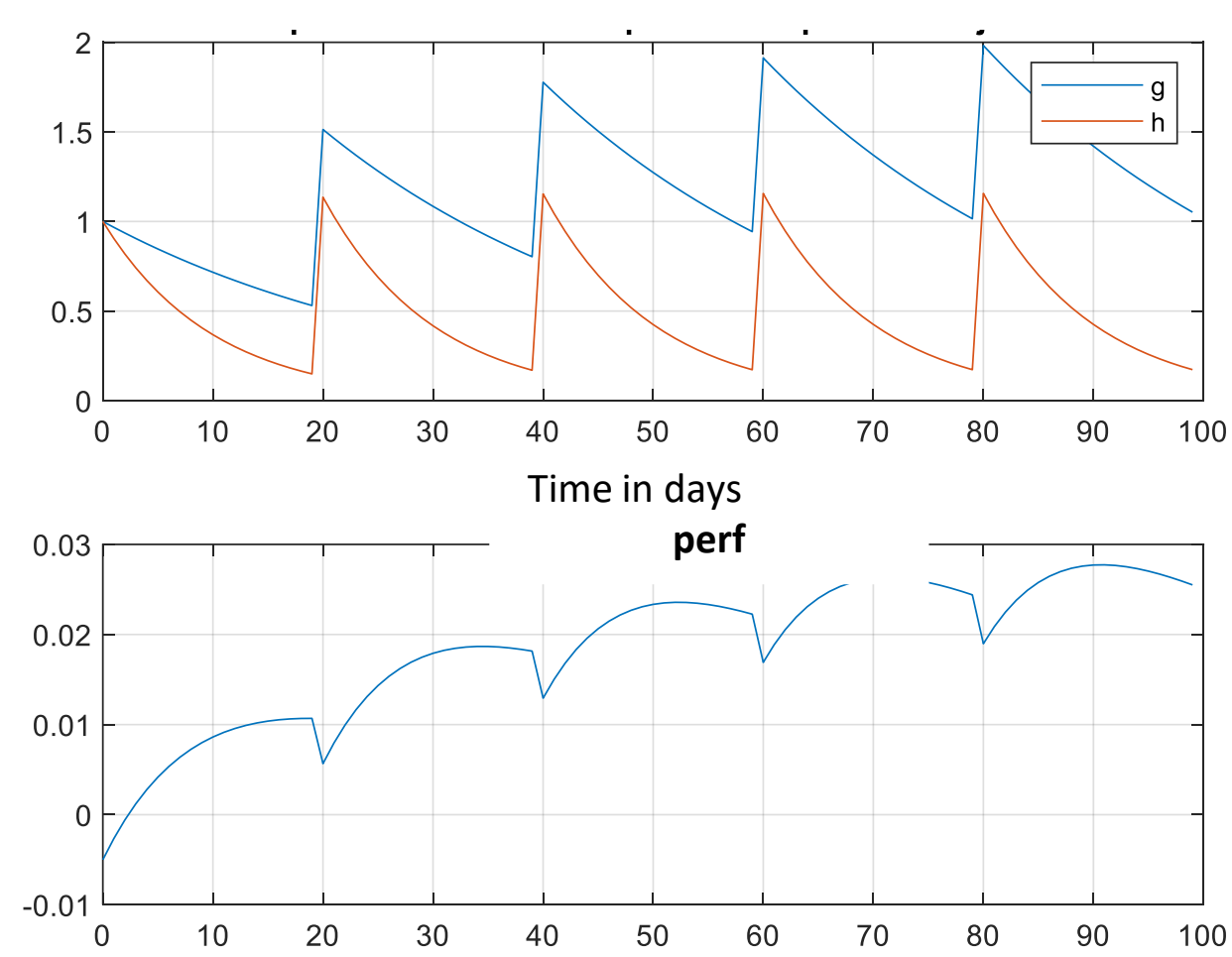

*Fig. 6: Response to a 20-day pulse train*

*With a 10-day pulse train (1 training session every 10 days):*

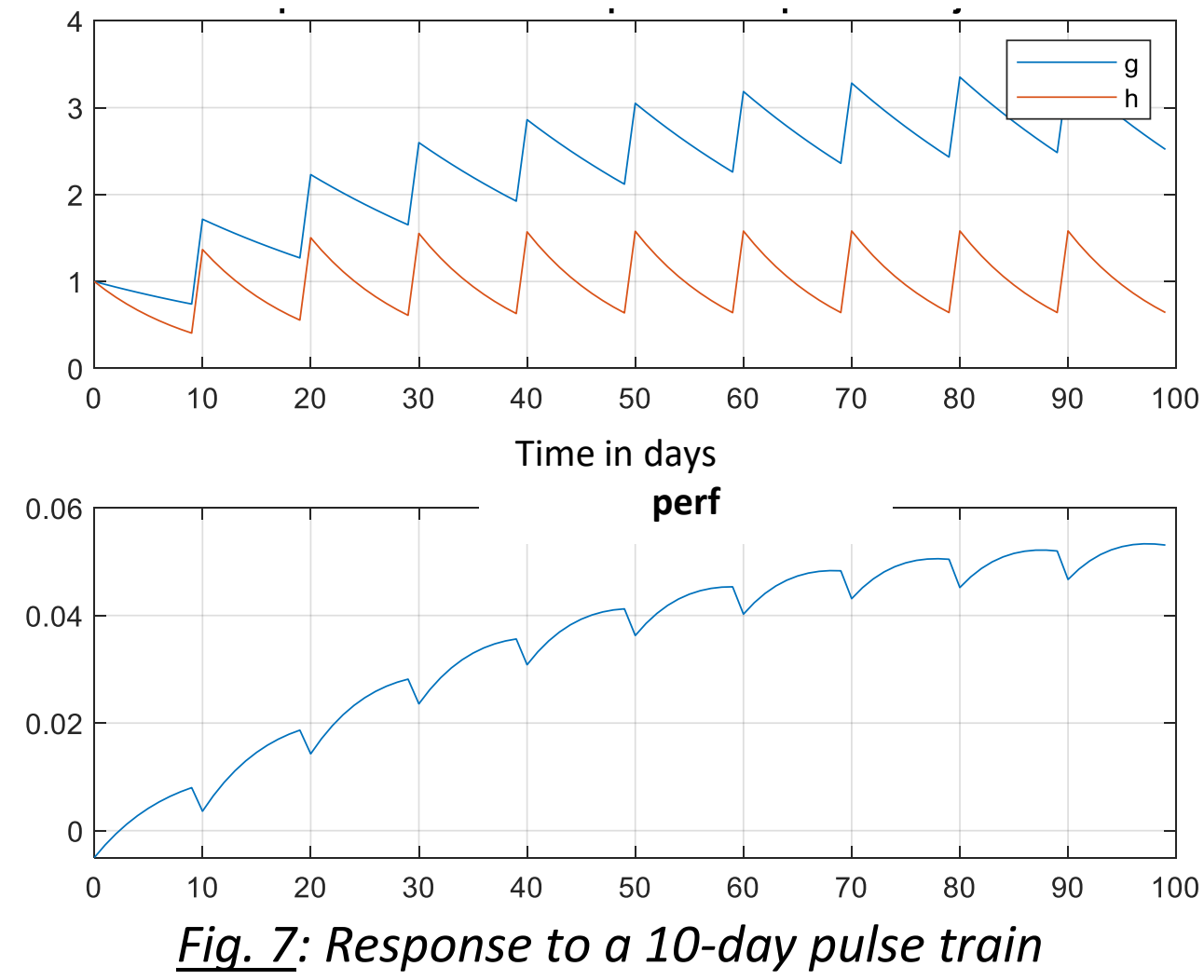

*Fig. 7: Response to a 10-day pulse train*

*With a pulse train of period 1 day (training every day):*

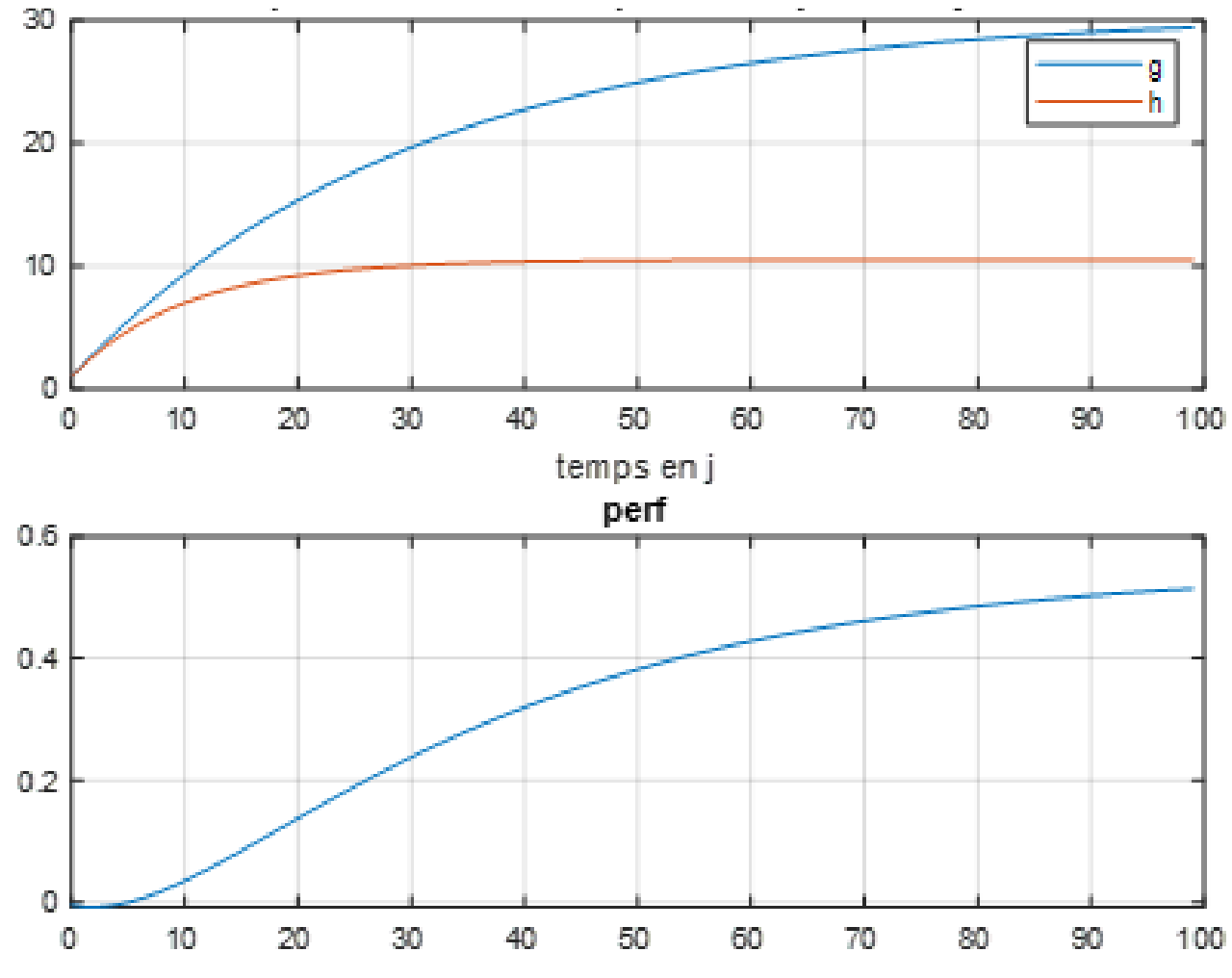

*Fig. 8: Response to a pulse train of period 1 day*

*The time constants of the system mean that when the frequency of training is daily, the jerks in response disappear completely.*

*Finally, let's look at the indicial response of the system, i.e. the response to a unit step ( $w(t)=1$ $if\ t\geq 0, 0\ otherwise$ ):*

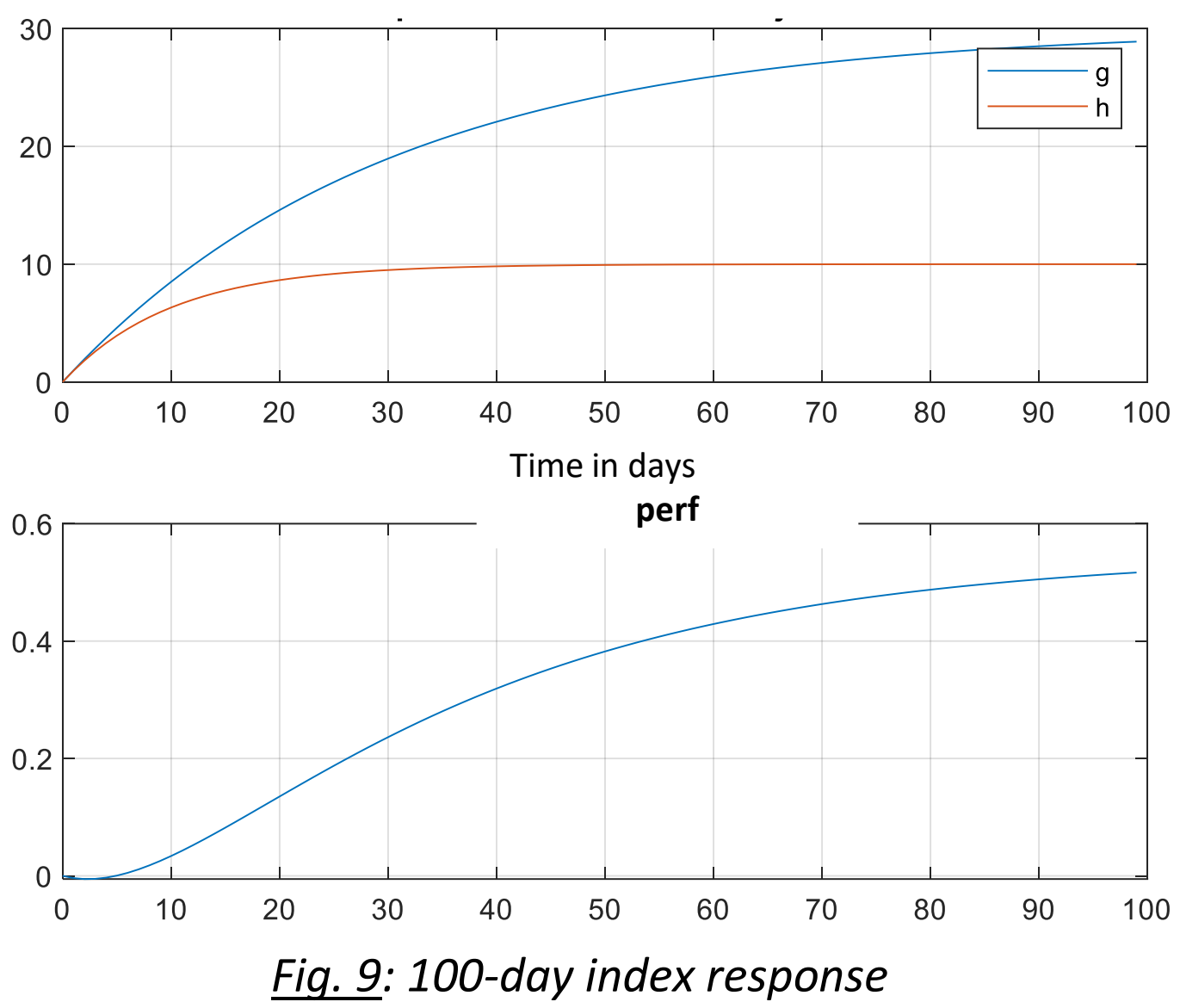

*Fig. 9: 100-day index response*

*There is no difference between the indicial response and the response to a pulse train corresponding to a daily training with the time constants of the fitness-fatigue model. For daily training, we can see $w(t)$ as a continuous signal whose amplitude can even be varied over the course of the day (alternating rest and effort times over the same day). Seeing $w(t)$ as a continuous signal seems to us to be a more realistic representation than reducing training to a pulse.*

We wanted to show by these calculations the genesis of the recurrence equations and convolution products that are used in the Banister model. The reader is thus given all the tools to imagine more complex models instead of the Banister model. This section aims to demystify the equations used in FFM models. We wanted to make explicit the mathematical formalisms that allow us to understand the convolution products used in FFM models. We have dedicated a point to the continuous/discrete duality of the representation of the load in the input of the FFM model: the idea is to show that when we will be able one day, thanks to the Internet of Things (IoT) for example, to observe the load as a continuous signal in quasi real time, the extension of the FFM model will be quite natural. Having explained the mathematical genesis of the FFM model through the state representation, it is also clear that any extension of Banister's original model to any continuous linear invariant system (CLI) is not a problem. We have recalled that the state formalism is very general and that it applies to any CLI system, whatever the number of inputs/outputs, whatever the order of the differential equations linking inputs and outputs, whatever the initial conditions, and that this formalism is reduced to two very simple equations: a first-order differential equation (admittedly matrix) and a linear output equation.

# 3. Identification of a linear dynamic model

We will now use the definition of the z-transform to identify the performance dynamical model.

We have just seen that for a linear model of a discretized dynamical system, we have:

$y(z)=H(z)u(z)=\frac{N(z)}{D(z)}u(z)$ (see equation(16)) where $N(z)$ and $D(z)$ are polynomials in $z$

or :

$$y(z)=\frac{b_1z^{-1}+b_2z^{-2}+b_3z^{-3}+..+b_mz^{-m}}{1+a_1z^{-1}+a_2z^{-2}+..+a_nz^{-n}}u(z) \tag{18}$$

This gives the recurrence equation in discrete time:

$$\begin{aligned}y(k+1)+a_1y(k)+a_2y(k-1)+..+a_ny(k-n+1)\\ =b_1u(k)+b_2u(k-1)+b_3u(k-2)+..+b_mu(k-m+1)\end{aligned}$$

Or:
$$\begin{aligned}y(k+1)=-a_1y(k)-a_2y(k-1)-..-a_ny(k-n+1)\\ +b_1u(k)+b_2u(k-1)+b_3u(k-2)+..+b_mu(k-m+1)\end{aligned}$$

The equation is available at $k-n+1\geq 0$ or: $k\geq n-1$ (19)

Let's imagine that we collect $N$ measurements, then we establish the following system of equations:

$$\begin{matrix} k=n-1 \\ k=n \\ \\ \\ k=n+N-1 \end{matrix}\underbrace{\begin{pmatrix} y(n) \\ y(n+1) \\ \\ \\ y(n+N) \end{pmatrix}}_{Y:N+1\times 1}=\underbrace{\begin{pmatrix} -y(n-1) & -y(n-2) & \dots -y(0) & u(n-1) & \dots u(n-1-m+1) \\ -y(n) & -y(n-1) & \dots -y(1) & u(n) & \dots\ u(n-m+1) \\ \dots & \dots & \dots\ \dots & \dots & \dots\ \dots \\ \dots & \dots & \dots\ \dots & \dots & \dots\ \dots \\ -y(n+N-1) & -y(n+N-2) & \dots -y(N) & u(n+N-1) & \dots\ u(n+N-m) \end{pmatrix}}_{H:(N+1)\times(n+m)}\underbrace{\begin{pmatrix} a_1 \\ \dots \\ a_n \\ b_1 \\ \dots \\ b_m \end{pmatrix}}_{\theta:(n+m)\times 1}$$

which we note: $Y=H\theta$ where $\theta$ is the vector of system parameters.

The measurements of *u* and *y* (they appear in $Y$ and $H$ ) are used to calculate the parameter vector $\theta$. $\theta$ is chosen to minimize the model error: $\|Y-H\theta\|$ .

The least squares criterion to be minimized is therefore defined: $J(\theta)=\frac{1}{2}(Y-H\theta)^T(Y-H\theta)$

Or by using the properties of the transpose:

$$2J(\theta) = (Y - H\theta)^T (Y - H\theta) = \left(Y^T - (H\theta)^T\right)(Y - H\theta)$$
$$= Y^T Y - Y^T H\theta - (H\theta)^T Y + (H\theta)^T H\theta = Y^T Y - Y^T H\theta - \theta^T H^T Y + (H\theta)^T H\theta$$

The derivative with respect to $\theta$ is calculated to find the minimum of $J$ :

$$2\frac{\partial J(\theta)}{\partial \theta} = 0 - H^T Y - H^T Y + 2H^T H\theta$$

$$\frac{\partial J(\theta)}{\partial \theta} = 0 \Rightarrow H^T H\theta = H^T Y \Rightarrow \theta = \left(H^T H\right)^{-1} H^T Y \qquad (20)$$

This is the analytical least squares solution.

The training of the dynamic model linking input and state is reduced to a least squares estimator and therefore requires a very reasonable amount of training data. In the case of the Banister model, there are only four parameters to identify (Busso *et al.*, 1997).

This algorithm allows very little data to be used to identify the FFM model. In any case, the amount of data required for parametric identification of the FFM model is nothing like the amount of data required for a machine learning approach, *i.e.* without assumptions about the relationships between inputs and outputs.

We have seen the abstraction that state representation allows and the efficiency of the associated mathematical tools, the ease of the identification stage of such models. We will now see the functionalities that state representation and linear automatics offer and that could be as many transcriptions of crucial questions in sports training sciences.

# 4. Control

A first question that the Baslister model could at least partially answer is the definition of the optimal training load: what input loads should be applied to optimise performance over a given range. In process control theory, this question is related to the control of dynamical systems. We can try to control a system by minimizing an energy criterion (here the fatigue of the athlete for example) or by imposing a forced dynamics of the system in closed loop. These are the two points of view that we will deal with in this section.

## 4.1. Optimal control

The initial system is represented by the state and observation equations: $\begin{cases} \dfrac{dX(t)}{dt} = AX(t) + BU(t) \\ Y(t) = CX(t) \end{cases}$

The optimal state feedback control of this system is described by the equations below. The idea is to adapt the input signal according to the state of the system. The input $U(t)$ becomes the control input of the system. A control law is implemented which is the resultant of a feedback from the state and a set point $Y^c(t)$, which is fixed a priori (we will take a unitary step in the simulations without losing generality):

$$U(t) = -FX(t) + GY^c(t)$$

The equations of the looped system then become:

$$\begin{cases} \frac{dX(t)}{dt} = (A - BF)X(t) + BGY^c(t) \\ Y(t) = CX(t) \end{cases} \quad (21)$$

$G$ is chosen so that $Y = Y^c$ in the static regime.

The diagram below summarises this modelling:

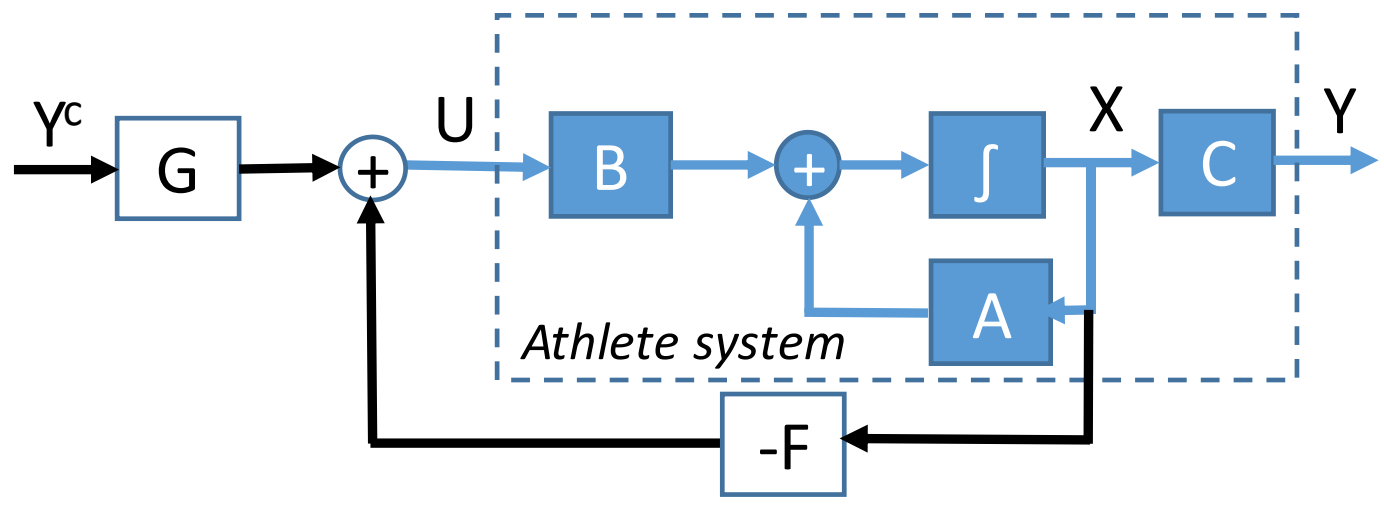

*Fig. 10: $Y^c$ the input of the system (the desired performance), U the control law (the load W) ,Y the output (the real performance) and X the state ([g; h] in the Banister model)*

The optimal quadratic control consists in searching for the matrix $F$ , which will minimize the following energy criterion $J$ :

$$J = \int_0^{+\infty} [X^t(t)QX(t) + U^t(t)RU(t)]dt$$

$$Q \geq 0, R > 0$$

The optimal state feedback control is given by the formula:

$$U(t) = -FX(t) \text{ with } F = \left(R + B^T PB\right)^{-1} B^T PA \quad (22)$$

where $P$ is the solution of the Ricatti equation:

$$P = Q + A^T \left(P - PB\left(R + B^T PB\right)^{-1} B^T P\right)A \quad (23)$$

Details of the calculations that lead to this result are provided in Annex 1 of this document.

***Example 8:** Let us illustrate the principle of optimal quadratic control with given values for Q and R in the framework of the Banister model. The larger the scalar R, the more important it is to take into account the load imposed on the athlete in the training strategy. The Q matrix can be seen as a weighting matrix in the simplest case where the Q matrix is diagonal. Thus, if Q is diagonal, the smaller $Q_{11}$ is compared to $Q_{22}$ , the more attention will be paid to not tiring the athlete in the training strategy; conversely, the smaller $Q_{22}$ is compared to $Q_{11}$, the more priority will be given to the dimension of fitness rather than that of fatigue during training (the athlete will not be hesitated to be tired).*

*The closed loop diagram is the following:*

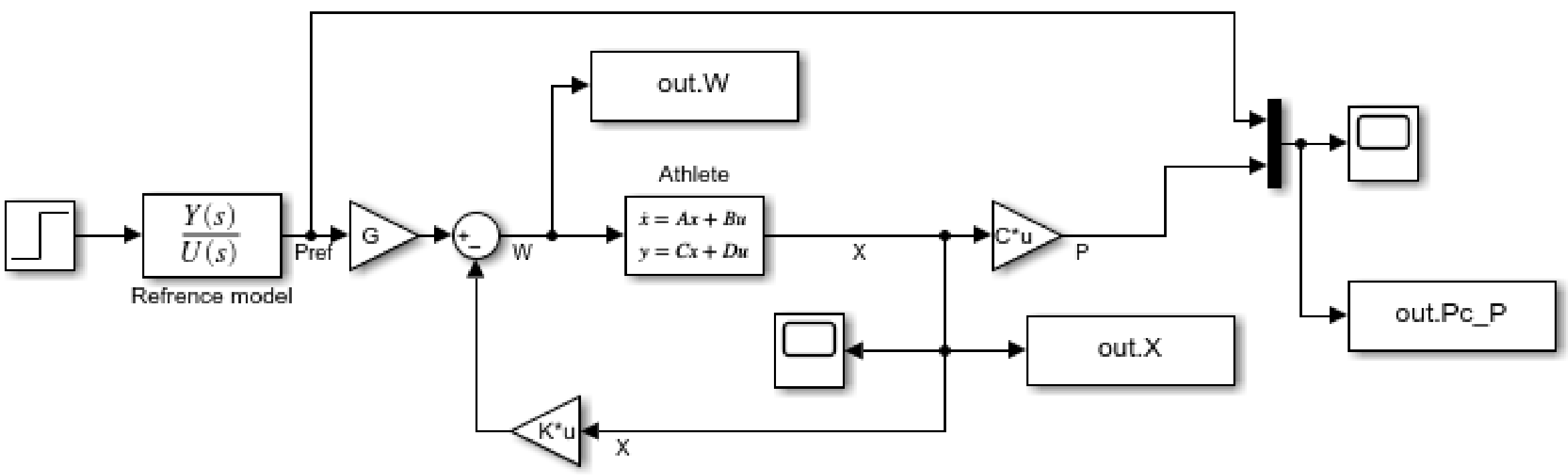

*Fig. 11: The closed-loop athlete system*

*Below is illustrated response to a unitary setpoint step filtered by a reference model (e.g., a simple 2 order system) to make the setpoint signal $P_{ref}$ smoother.*

*Quadratic control, indicial response* $R = 10$ and $Q = \begin{pmatrix} 1 & 0 \\ 0 & 100 \end{pmatrix}$

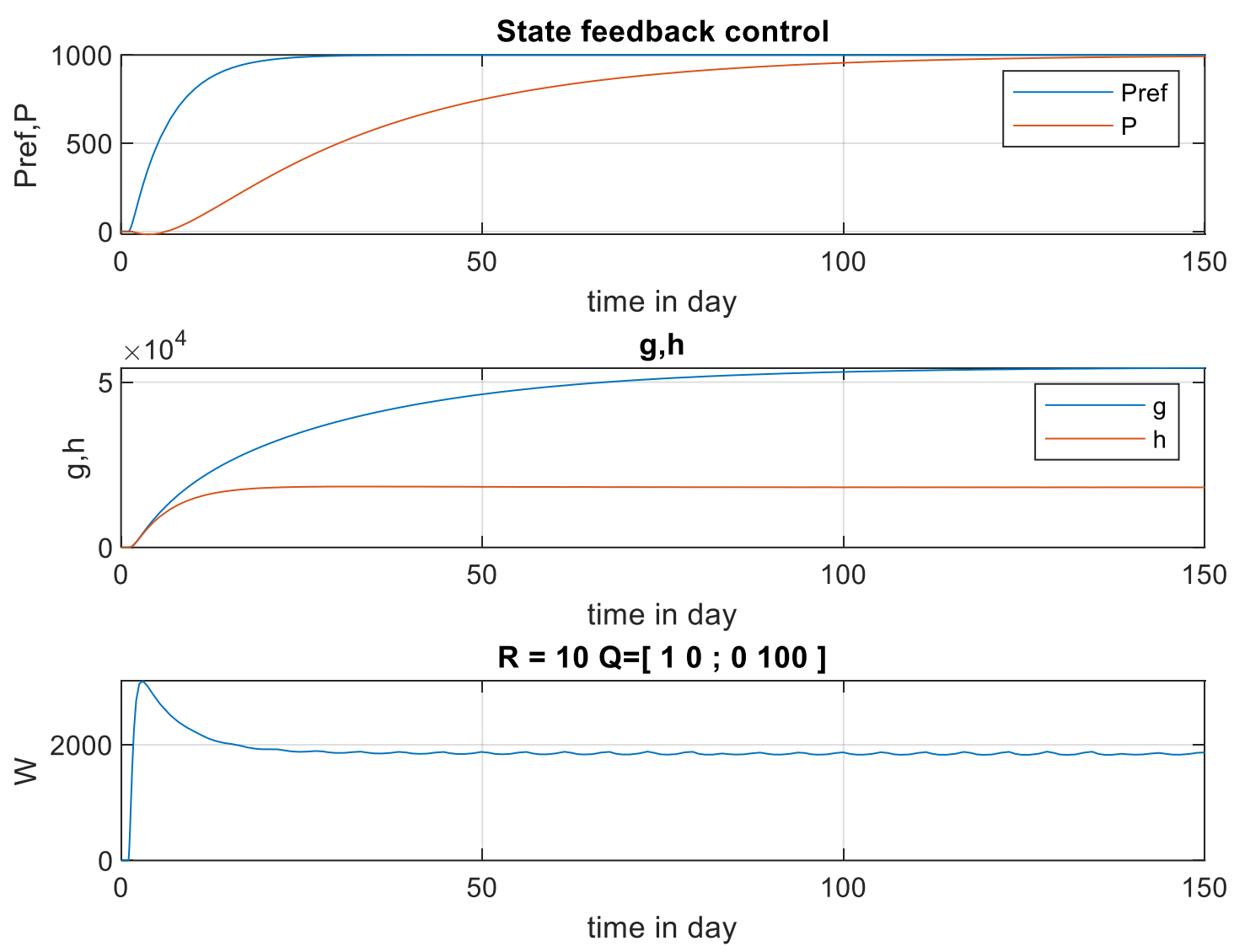

*Fig. 12: With these values of R and Q, the importance given to the load, R, is consequent, and the weight given to fatigue ($Q_{22}$) is very great compared to the weight given to fitness ($Q_{11}$). In other words, choosing these parameters R and Q corresponds to a training strategy where the main thing is not to tire the athlete. After 70 days of training, however, the performance has still not reached its desired static value of 1! These values could of course be refined, in particular by decreasing R, in order to have a faster response, but the aim here was simply to illustrate a training strategy whose objective is to save and protect the athlete.*

In conclusion, when we have a linear dynamic model of the system, we can analytically define the state feedback *F* to be applied and then define analytically the control law that will minimise a given energy criterion. Optimal control replaces in this case a reinforcement learning algorithm that would require a substantial learning base to identify the optimal load to apply according to the state of the athlete.

## 4.2. Control by state feedback and pole placement

As before, a control law is implemented which is the result of a feedback from the state and a setpoint $Y^c(t)$ (the drive load) which is fixed a priori: $U(t) = -FX(t) + GY^c(t)$ . The equations of the closed-loop system are therefore as in the previous subsection:
$$\begin{cases} \dfrac{dX(t)}{dt} = (A - BF)X(t) + BGY^c(t) \\ Y(t) = CX(t) \end{cases}$$

and $G$ is chosen so that $Y = Y^c$ in static regime ( $G = pinv(C(BF - A)^{-1}B)$ ).

The difference with the approach known as optimal quadratic control lies in the strategy of choosing $F$ : the state feedback is no longer chosen in order to minimize this or that energy criterion as previously, but in order to choose the dynamics of the looped system. The dynamics of the closed loop system is $(A - BF, BG, C)$ where $(A, B, C)$ is the dynamics of the initial system . In order to choose the dynamics of the looped system, we are therefore interested in the poles or eigenvalues of $A - BF$ . These eigenvalues must already be chosen so that the looped system is asymptotically stable (negative real parts). Eigenvalues should also be chosen that are pure reals in order to avoid oscillations. For a system of dimension $n$ , we therefore have $n$ poles to be fixed $[p_1, p_2, .., p_n]$ which are chosen for instance such as $p_1 = -\dfrac{1}{\tau_d}$ and $\forall p_i, i \neq 1, |p_i| \gg \dfrac{1}{\tau_d}$ so that the looped system has a dominant transient response which corresponds to the response of a first order time constant $\tau_d$ since the time constants corresponding to the other poles are negligible compared to $\tau$ (quasi-instantaneous response on the scale of $\tau_d$ ). This is called state feedback control by choosing the poles of the closed loop system.

**Example 9:** *The idea is to a priori choose the eigenvalues $\lambda_1$ and $\lambda_2$ of $A - BF$ in the 2-dimensional Banister's model and then compute the feedback control $F$ that provides the expected dynamics of the closed loop system. This makes it possible to control the date by which a performance must be achieved.*

$$A - BF = \begin{pmatrix} -\dfrac{1}{\tau_g} & 0 \\ 0 & -\dfrac{1}{\tau_h} \end{pmatrix} - \begin{pmatrix} 1 \\ 1 \end{pmatrix} \begin{pmatrix} f_1 & f_2 \end{pmatrix} = \begin{pmatrix} -\dfrac{1}{\tau_g} & 0 \\ 0 & -\dfrac{1}{\tau_h} \end{pmatrix} - \begin{pmatrix} f_1 & f_2 \\ f_1 & f_2 \end{pmatrix} \text{ with } F = \begin{pmatrix} f_1 & f_2 \end{pmatrix}$$

$$A - BF = \begin{pmatrix} -\dfrac{1}{\tau_g} - f_1 & -f_2 \\ -f_1 & -\dfrac{1}{\tau_h} - f_2 \end{pmatrix}$$

*The eigenvalues $\lambda_1$ and $\lambda_2$ of $A - BF$ are such that:*

$$Tr(A - BF) = -\left( \frac{1}{\tau_g} + f_1 + \frac{1}{\tau_h} + f_2 \right) = \lambda_1 + \lambda_2$$

$$\det\left(A-BF\right)=\left(\frac{1}{\tau_g}+f_1\right)\left(\frac{1}{\tau_h}+f_2\right)-f_1 f_2=\lambda_1\lambda_2$$

*This leads to:*

$$f_1=\frac{\lambda_1\lambda_2+\frac{1}{\tau_g}\left(\frac{1}{\tau_g}+\lambda_1+\lambda_2\right)}{\left(\frac{1}{\tau_h}-\frac{1}{\tau_g}\right)}=\frac{\tau_g\tau_h\left(\lambda_1\lambda_2+\frac{1}{\tau_g}\left(\frac{1}{\tau_g}+\lambda_1+\lambda_2\right)\right)}{\left(\tau_g-\tau_h\right)} \tag{24}$$

$$f_2=-\lambda_1-\lambda_2-\frac{1}{\tau_g}-\frac{1}{\tau_h}-\frac{\tau_g\tau_h\left(\lambda_1\lambda_2+\frac{1}{\tau_g}\left(\frac{1}{\tau_g}+\lambda_1+\lambda_2\right)\right)}{\left(\tau_g-\tau_h\right)} \tag{25}$$

*The state feedback matrix $F$ is calculated by choosing the eigenvalues $\lambda_1,\lambda_2<0$ (asymptotically stable system) and $|\lambda_1|\gg|\lambda_2|$ (dominant transient response) in order to obtain the desired dynamics for the looped system.*

**Example 10:** *The simulation is processed with $\tau_d=6\,days$ as the closed-loop time constant, and the setpoint is simply modified with a second-order reference model (same filter as in example 8). This experimentation makes it possible to have a "softer" signal at the input in order to avoid too violent initial load increases (horizontal tangent at the origin).*

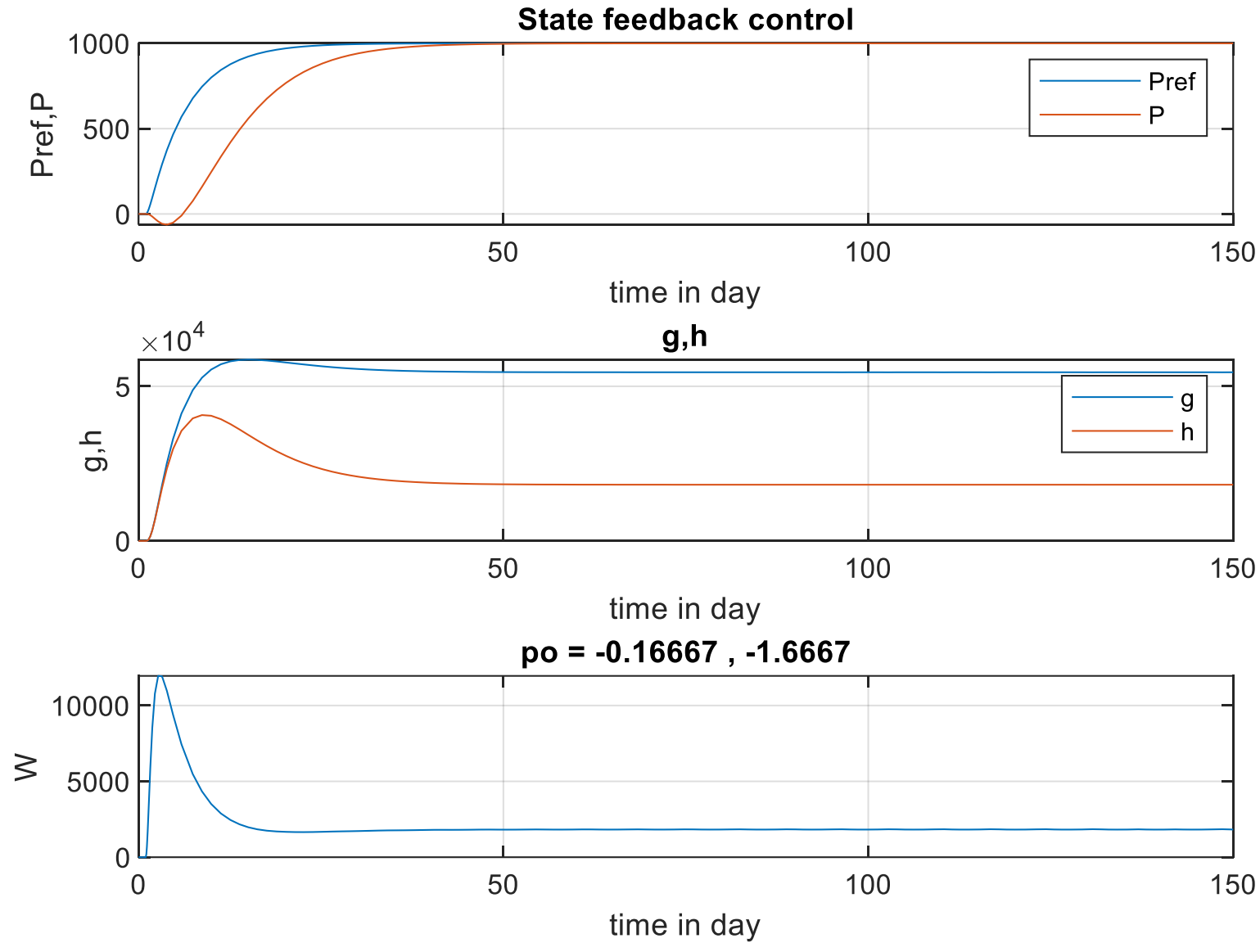

*Fig. 13: Response with a pole placement approach $\tau_d=6\,days$*

*The reference model makes the effort rise slower. The change in set point is smoothed by the reference model. This closed-loop system appears particularly relevant since it has a very interesting response time while limiting the amplitude of the initial load. Contrary to the study of open-loop loading, where only the amplitude of the load was controlled without any guarantee of the response time, this time the dynamics of the performance are controlled while being able to define a template to follow. In*

*practice, this model makes it possible to determine the control law* $w(t)$ *which takes into account the state of the athlete (i.e.* $w(t) = G\,\mathrm{P}_{ref}(t) - K\begin{bmatrix} g(t) \\ h(t) \end{bmatrix}$*), i.e. the effective load to be applied to the athlete, in order to follow the expected performance.*

## 5. The observer

In general, the state of a system is not completely observed (due to a lack of sensors for technical or financial reasons), and an estimator is sought. This estimator corresponds to the diagram of an observer $K$ as shown in Figure 14. The aim is to obtain an estimate of the state, *i.e.* the fitness variables *g* and the fatigue variables *h* in the Banister model, while only the output *P is* measured. Remember that most studies in the literature assume that the state variables are zero at $t = 0$. It is because only variations around an equilibrium phase are studied where state variables are fixed but not null (except after a long rest period).

**Example 11:** *Thus, the problems of imputing the load perceived by the athlete during a training session could be brought closer to the construction of a state observer. The estimator could be used to estimate fatigue and fitness when the Rating of Perceived Exertion (RPE) is not available at each training session. An interesting manipulation could also be to compare the state estimated by the observer with the data from the RPE assessment: if the estimator proves reliable then the "RPE" could be systematically calculated after each training session. If the fitness variables g and fatigue variables h can be simulated with the previous equations, the observer allows the model to be "recalibrated" with the available observations, in this case the performance, for a more precise estimation of the state variables or the detection of disturbances. In particular, it allows the initial values of the state variables to be eliminated.*

The result we are interested in is the following:

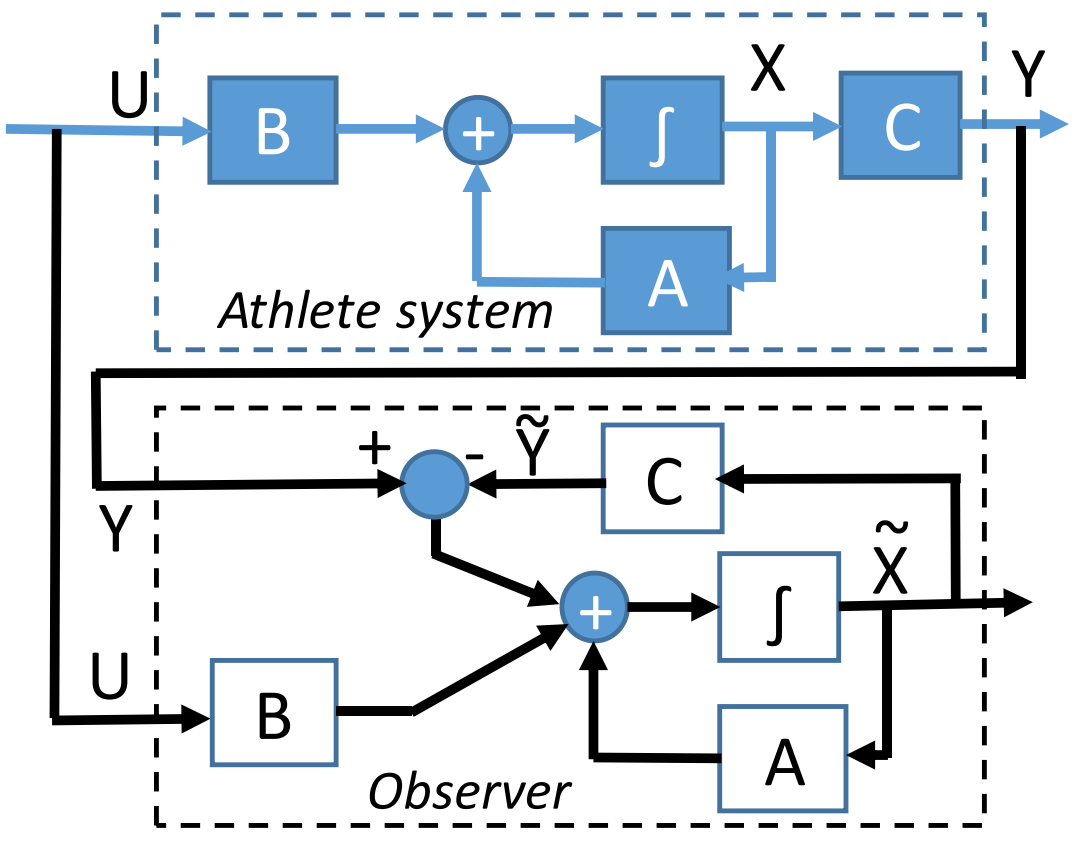

*Fig. 14: The observer scheme*

Let us detail the calculations in the framework of discrete modelling. In the discretized representation, we have seen that the state equations of the system are:

System: $\begin{cases} X_{k+1} = \bar{A}X_k + \bar{B}U_k \\ Y_k = \bar{C}X_k \end{cases}$ and for the observer of Fig. 14:

Observer: $\begin{cases} \widehat{X}_{k+1} = \overline{A}\widehat{X}_k + \overline{B}U_k + K\left(Y_k - \widehat{Y}_k\right) = \overline{A}\widehat{X}_k + \overline{B}U_k + K\overline{C}\left(X_k - \widehat{X}_k\right) \\ \widehat{Y}_k = \overline{C}\widehat{X}_k \end{cases}$ (26)

$K\overline{C}\left(X_k - \widehat{X}_k\right) = K\left(Y_k - \widehat{Y}_k\right)$ is the innovation factor, it allows the model to be adjusted to the observations. The drift of the model is thus controlled. Now let us introduce the estimation error $E_k = X_k - \widehat{X}_k$ and establish the recurrence equation it verifies.

$$\begin{aligned} &\widehat{X}_{k+1} = \overline{A}\widehat{X}_k + \overline{B}U_k + K\left(Y_k - \widehat{Y}_k\right) = \overline{A}\widehat{X}_k + \overline{B}U_k + K\overline{C}\left(X_k - \widehat{X}_k\right) \\ &\widehat{Y}_k = \overline{C}\widehat{X}_k \\ &\Rightarrow E_{k+1} = X_{k+1} - \widehat{X}_{k+1} = \overline{A}\left(X_k - \widehat{X}_k\right) - K\overline{C}\left(X_k - \widehat{X}_k\right) = \left(\overline{A} - K\overline{C}\right)E_k \end{aligned} \quad (27)$$

This gives the estimate :

$$\widehat{X}_{k+1} = \left(\overline{A} - K\overline{C}\right)\widehat{X}_k + \overline{B}U_k + K\overline{C}X_k = \left(\overline{A} - K\overline{C}\right)\widehat{X}_k + \overline{B}U_k + KY_k = \left(\overline{A} - K\overline{C}\right)\widehat{X}_k + \left(\overline{B} \quad K\right)\begin{pmatrix} U_k \\ Y_k \end{pmatrix} \quad (28)$$

Note that if we are interested in the asymptotic stability in this discrete representation, the eigenvalues of $\overline{A} - K\overline{C}$ must be of modulus strictly less than 1 (reason for a convergent geometric sequence). The calculations below illustrate how an estimate of the state is obtained from the measured output *P* and the input $\omega$.

**Example 12:** *Let us construct an observer for the Banister model, and take the continuous time model to illustrate the two perfectly equivalent views. The state equation for the observer* $X = \begin{pmatrix} \tilde{g}(t) \\ \tilde{h}(t) \end{pmatrix}$ *, where the state estimate is corrected using the output measurement, is:*

$$\begin{pmatrix} \dot{\tilde{g}}(t) \\ \dot{\tilde{h}}(t) \end{pmatrix} = A\begin{pmatrix} \tilde{g}(t) \\ \tilde{h}(t) \end{pmatrix} + B\omega(t) + K\left(P(t) - C.\begin{pmatrix} \tilde{g}(t) \\ \tilde{h}(t) \end{pmatrix}\right)$$

$$\begin{pmatrix} \dot{g}(t) - \dot{\tilde{g}}(t) \\ \dot{h}(t) - \dot{\tilde{h}}(t) \end{pmatrix} = \left(\begin{pmatrix} -\frac{1}{\tau_g} & 0 \\ 0 & -\frac{1}{\tau_h} \end{pmatrix} - \begin{pmatrix} K_1 \\ K_2 \end{pmatrix}\begin{pmatrix} k_g & k_h \end{pmatrix}\right)\begin{pmatrix} \left(g - \tilde{g}\right)(t) \\ \left(h - \tilde{h}\right)(t) \end{pmatrix}$$

*By introducing the differential equation of the estimation error:*

$$\dot{E}(t)=\left(\begin{pmatrix}-\frac{1}{\tau_g} & 0\\ 0 & -\frac{1}{\tau_h}\end{pmatrix}-\begin{pmatrix}K_1\\ K_2\end{pmatrix}\begin{pmatrix}k_g & k_h\end{pmatrix}\right)E(t)=-\begin{pmatrix}\frac{1}{\tau_g}+K_1k_g & K_1k_h\\ K_2k_g & \frac{1}{\tau_h}+K_2k_h\end{pmatrix}E(t)$$

$$\dot{E}(t)=\left(A-KC\right)E\left(t\right)\text{ with } A-KC=-\begin{pmatrix}\frac{1}{\tau_g}+K_1k_g & K_1k_h\\ K_2k_g & \frac{1}{\tau_h}+K_2k_h\end{pmatrix} \quad (29)$$

*In general, we want the error $E(t)$ to converge "quickly" to zero (the measurement and the estimate match as quickly as possible) so that the estimator can be used from the first measurements. For this, all eigenvalues of $A-KC$ must be negative real so that $A-KC$ is asymptotically stable.*

Calculation of eigenvalues:

$$\begin{vmatrix}\frac{1}{\tau_g}+K_1k_g-\lambda & K_1k_h\\ K_2k_g & \frac{1}{\tau_h}+K_2k_h-\lambda\end{vmatrix}=\lambda^2-\lambda\left(\frac{1}{\tau_g}+K_1k_g+\frac{1}{\tau_h}+K_2k_h\right)+\left(\frac{1}{\tau_g\tau_h}+\frac{K_1k_g}{\tau_h}+\frac{K_2k_h}{\tau_g}\right)$$

$$=\lambda^2-a\lambda+b$$

$$\Delta=a^2-4b$$

$$\lambda=\frac{a\pm\sqrt{\Delta}}{2}\ si\ \Delta>0\ et\ \lambda=\frac{a\pm i\sqrt{|\Delta|}}{2}\ si\ \Delta<0$$

*The asymptotic stability conditions are:* $\left(\Delta>0\right)\wedge\left(a+\sqrt{\Delta}<0\right)$ *or* $\left(\Delta<0\right)\wedge\left(a<0\right)$ .

**Example 13:**

*The previous section proposed a control loop that assumes that the initial state variables values $g\left(t\right)$ and $h\left(t\right)$ are known (which is generally not the case since only the ouput $P\left(t\right)$ is known ). Controlling with an observer allows similar results to be obtained without these assumptions. In practice, the physical trainer does not have to worry about the athlete's state of fatigue or fitness, as the observer makes it possible to estimate the values reliably after a response time fixed by the choice of parameters in the observer's gain matrix $K$ (these parameters fix the observer's convergence dynamics).*

*A reference model of second-order response without overshoot is kept to ensure graduality in the evolution of the load as in example 10 and it is desired that the closed-loop system responds as a first-order time constant $\tau_d=6days$ .*

*We have moreover simulated a "perturbation event". This event modifies the value of the performance. This disturbance takes the form of a step (time 100) in the following simulation: we have imagined that the athlete had a drop in performance that set in at a certain point in the season and persisted until*

*the end of the season, but we could have envisaged any type of disturbance, whether ephemeral or lasting.*

*We also introduced a ramp as a setpoint signal in order to consider an objective of linear progression of the athlete's performance over time. As a consequence, to guarantee the precision performance, the control input is no more* $w(t)$ *but* $\dfrac{dw(t)}{dt}$.

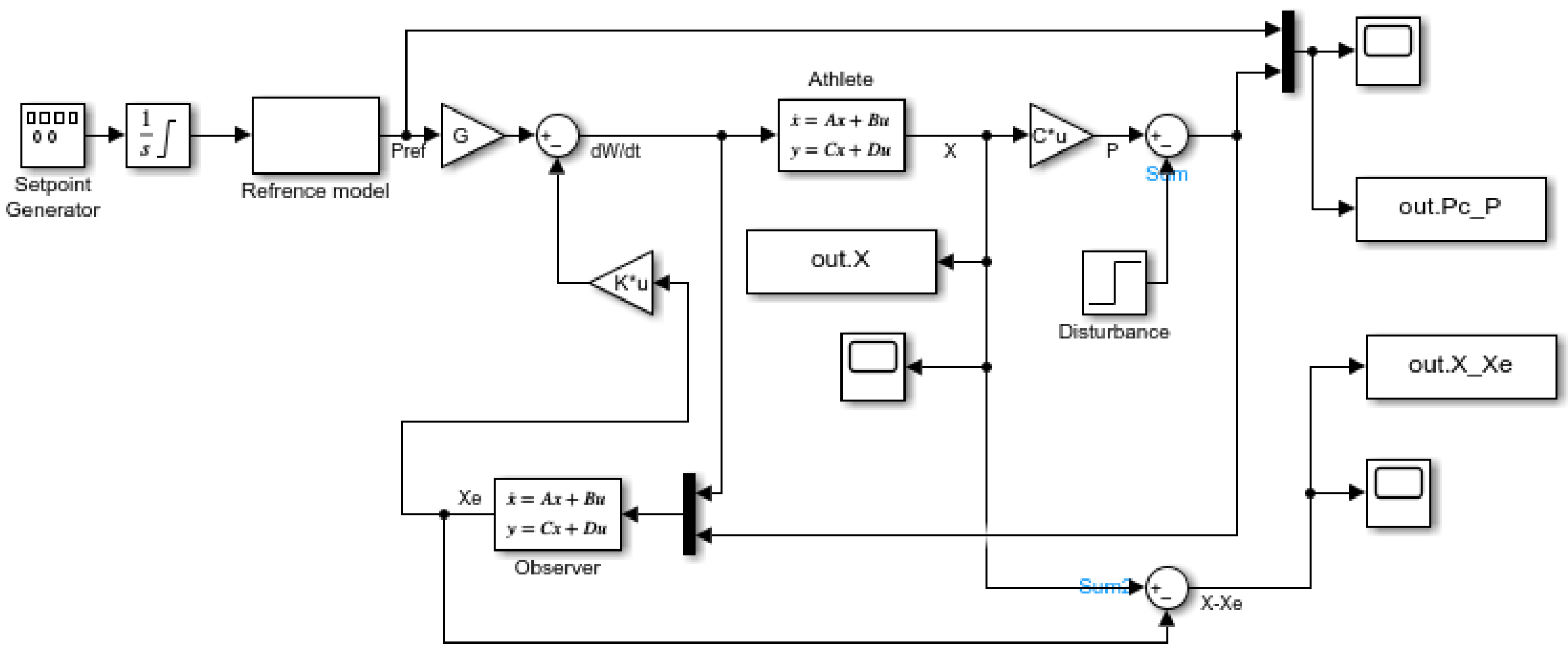

*Fig. 15: Matlab schematic of the controlled system with state observer*

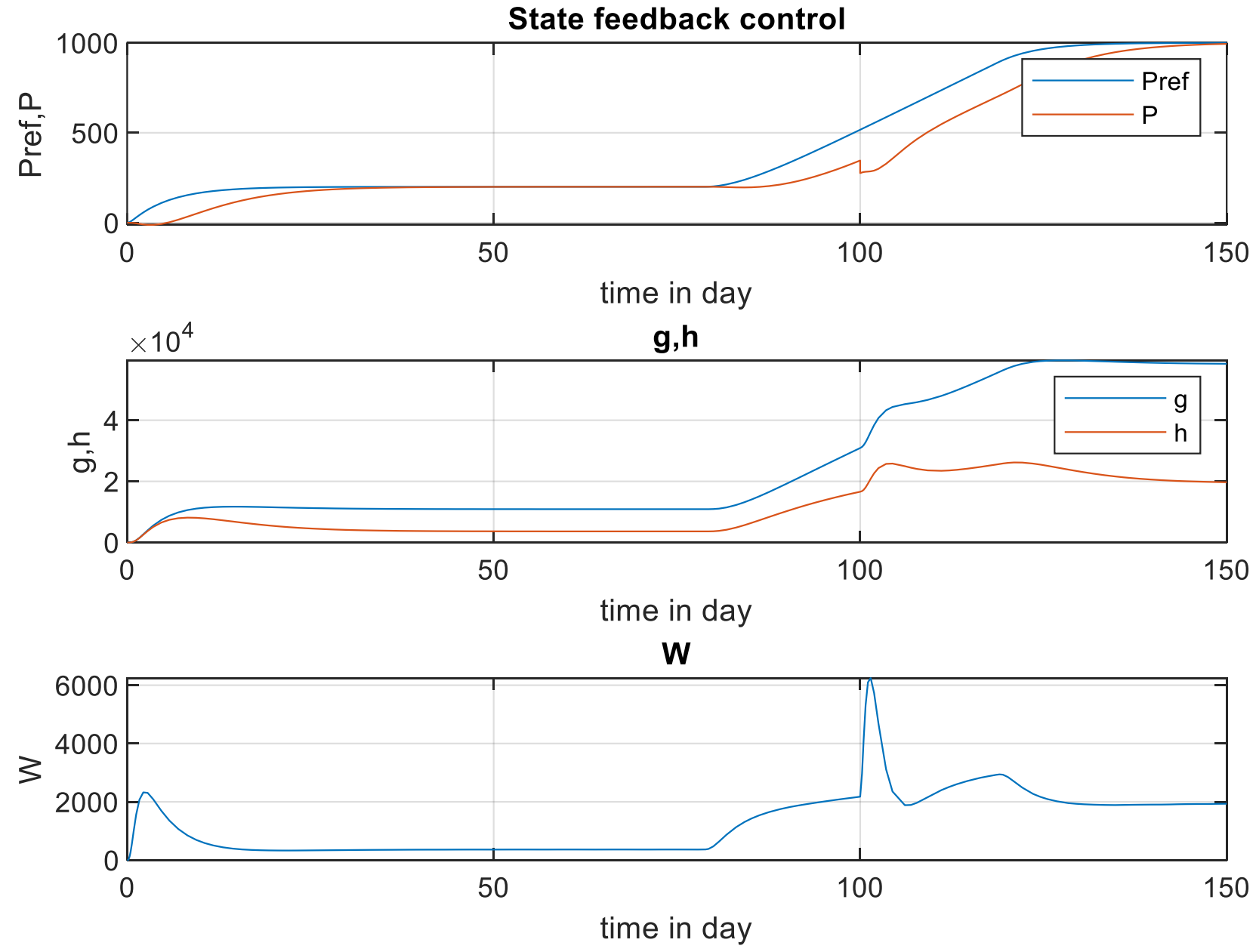

*Fig. 16: Response of the servo system with a conventional observer to a reference performance curve*

*It is the estimate of the state that enters the feedback loop, since* $g(t)$ *and* $h(t)$ *are not known in practice. It can be seen that the observer plays its role well (Fig.17) and the real performance* $P(t)$ *pursues the reference performance* $Pref(t)$ *with zero static error. At instant 100, the performance perturbation is detected and corrected with an increase on the load* $w(t)$.

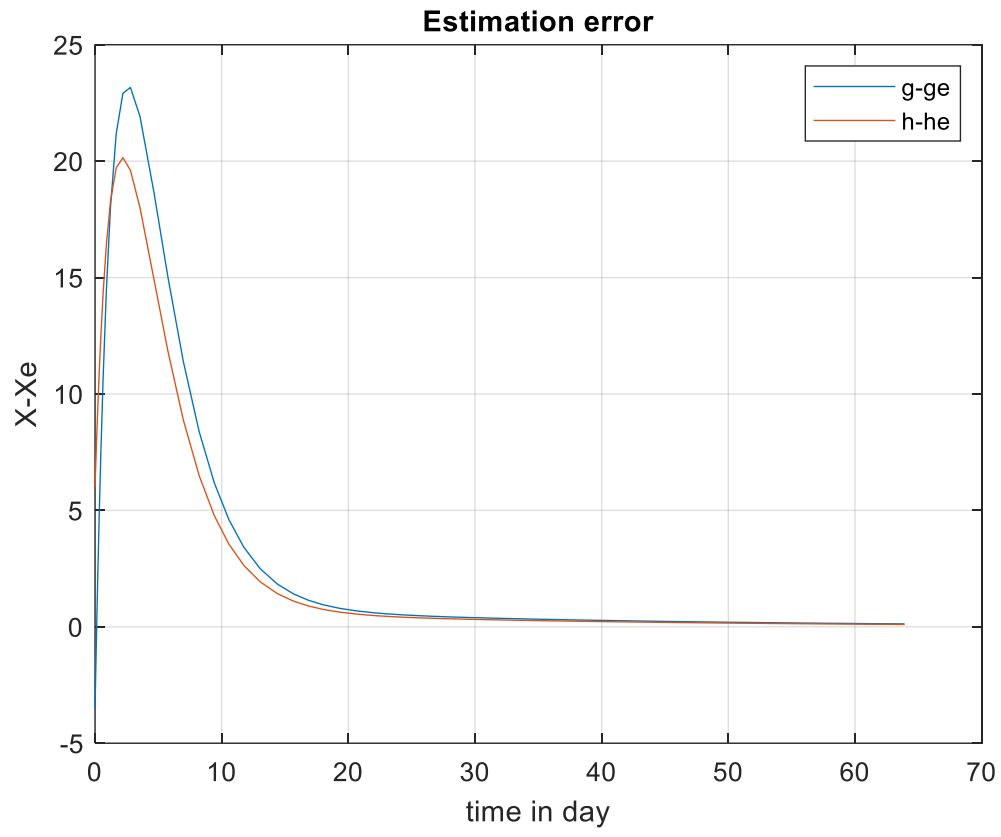

*Fig. 17: The estimation error converges to zero, the observer plays its role well.*

# 6. Model-based diagnostics and analytical redundancies

The next use we wish to mention of the dynamic behaviour model is the diagnosis of system malfunctions. The idea is to predict a system malfunction before the consequences are manifest and directly observable. In our example, the model will make it possible to highlight an inconsistency between the measurements and the inputs of the system well before the behaviour of the athlete becomes alarming for the human observer.

Model-based diagnosis means that one is able to compare process and model. This implies that a symptom in this type of system is a deviation, called a residual, between the process and model behaviour.

The measured variables of a system are often linked by a set of relationships that define the functional behaviour of the system. The purpose of static redundancy is to find the algebraic relationships between the instantaneous values of the measurements. In the linear case, the measurement vector $Y$ is related to the state vector $X$ by:

$$Y(k) = CX(k) + \varepsilon \tag{30}$$

where $\varepsilon$ is the error vector and $C$ the observation matrix.

If there is a parity matrix $Y$ that verifies $VC = 0$ , then the measures are said to verify the parity relations obtained by elimination of $x : P = VY = V\varepsilon$ .

Static redundancy therefore exploits the relationships between instantaneous or stabilized values of observations of a physical system. This means that the output of one sensor can be reconstructed from the other sensors. The appearance of an abnormal deviation of the residual is indicative of a failure.

If a dynamic model of the system is available, this residuals generation can be generalized as follows:

$$X(k+1) = AX(k) + BU(k)$$

$$Y(k) = CX(k)$$

$$\begin{pmatrix} Y(k) \\ Y(k+1) \\ Y(k+2) \\ \\ Y(k+r) \end{pmatrix} = \begin{pmatrix} C \\ CA \\ CA^2 \\ \\ CA^r \end{pmatrix} X(k) + \begin{pmatrix} 0 & 0 & 0..0 & 0 \\ CB & 0 & 0..0 & 0 \\ CAB & CB & 0..0 & 0 \\ \\ CA^{r-2}B & CA^{r-3}B & .. & 0 \\ CA^{r-1}B & CA^{r-2}B & .. & CB \end{pmatrix} \begin{pmatrix} U(k) \\ U(k+1) \\ \\ \\ U(k+r-1) \end{pmatrix} \tag{31}$$

or in the presence of noise and simplifying the notations, (19) becomes:

$$Y(k,r) = \mathbf{C}(r).X(k) + \mathbf{B}(r).U(k,r) + \varepsilon(r) \tag{32}$$

The redundancy relations are obtained by multiplying the system by the parity matrix $\Omega$ as $\Omega.\mathbf{C}(r) = 0$ with $\mathbf{C}(r)$ the generalized observability matrix.

The generalized parity equations (which provide the residuals) then have the expression:

$$\underbrace{P}_{r\times 1} = \underbrace{\Omega}_{r\times r} \left[ \underbrace{Y(k,r)}_{r\times 1} - \underbrace{\mathbf{B}(r)}_{r\times r} . \underbrace{U(k,r)}_{r\times 1} \right] = \underbrace{\Omega\varepsilon(r)}_{r\times 1} \tag{33}$$

(note: we have $\underbrace{P}_{r\times 1} = \underbrace{\Omega}_{r\times r} \underbrace{\mathbf{C}(r)}_{r\times 2} \underbrace{x(k)}_{2\times 1}$) (in our case the dimension of the state is 2, *h* and *g*).

Note that not all these equations are necessarily independent, especially if *r* is large.

**Example 14:** *Returning to the Banister model:*

$$\begin{cases} \dfrac{dg(t)}{dt} = \omega(t) - \dfrac{1}{\tau_g}.g(t) \\ \dfrac{dh(t)}{dt} = \omega(t) - \dfrac{1}{\tau_h}.h(t) \end{cases} \Leftrightarrow \begin{pmatrix} \dot{g}(t) \\ \dot{h}(t) \end{pmatrix} = \underbrace{\begin{pmatrix} -\dfrac{1}{\tau_g} & 0 \\ 0 & -\dfrac{1}{\tau_h} \end{pmatrix}}_{A} \begin{pmatrix} g(t) \\ h(t) \end{pmatrix} + \underbrace{\begin{pmatrix} 1 \\ 1 \end{pmatrix}}_{B} \omega(t)$$

$$\Delta P = k_g.g(t) - k_h.h(t) = \begin{pmatrix} k_g & -k_h \end{pmatrix} \begin{pmatrix} g(t) \\ h(t) \end{pmatrix} = CX(t)$$

*Then,*

$$C = \begin{pmatrix} k_g & -k_h \end{pmatrix}$$

$$CA = \begin{pmatrix} k_g & -k_h \end{pmatrix} \begin{pmatrix} -\dfrac{1}{\tau_g} & 0 \\ 0 & -\dfrac{1}{\tau_h} \end{pmatrix} = \begin{pmatrix} -\dfrac{k_g}{\tau_g} & \dfrac{k_h}{\tau_h} \end{pmatrix}$$

*More generally by associativity:*

$$CA^r = \begin{pmatrix} k_g & -k_h \end{pmatrix} \begin{pmatrix} \left(-\frac{1}{\tau_g}\right)^r & 0 \\ 0 & \left(-\frac{1}{\tau_h}\right)^r \end{pmatrix} = \begin{pmatrix} \frac{(-1)^r k_g}{\tau_g{}^r} & \frac{(-1)^{r+1} k_h}{\tau_h{}^r} \end{pmatrix}$$

$$CB = \begin{pmatrix} k_g & -k_h \end{pmatrix} \begin{pmatrix} 1 \\ 1 \end{pmatrix} = k_g - k_h$$

$$CAB = \begin{pmatrix} k_g & -k_h \end{pmatrix} \begin{pmatrix} -\frac{1}{\tau_g} & 0 \\ 0 & -\frac{1}{\tau_h} \end{pmatrix} \begin{pmatrix} 1 \\ 1 \end{pmatrix} = \begin{pmatrix} k_g & -k_h \end{pmatrix} \begin{pmatrix} -\frac{1}{\tau_g} \\ -\frac{1}{\tau_h} \end{pmatrix} = -\frac{k_g}{\tau_g} + \frac{k_h}{\tau_h}$$

$$CA^r B = \begin{pmatrix} k_g & -k_h \end{pmatrix} \begin{pmatrix} \left(-\frac{1}{\tau_g}\right)^r & 0 \\ 0 & \left(-\frac{1}{\tau_h}\right)^r \end{pmatrix} \begin{pmatrix} 1 \\ 1 \end{pmatrix} = \begin{pmatrix} k_g & -k_h \end{pmatrix} \begin{pmatrix} \left(-\frac{1}{\tau_g}\right)^r \\ \left(-\frac{1}{\tau_h}\right)^r \end{pmatrix} = \begin{pmatrix} \frac{(-1)^r k_g}{\tau_g{}^r} + \frac{(-1)^{r+1} k_h}{\tau_h{}^r} \end{pmatrix}$$

*Depending on the values of r, the generated redundancy equations are not necessarily independent. The basic principle is to start a residual generation with small r and increment.*

*We are looking for* $\Omega(r \times r)$ *such that:*

$$\Omega \begin{pmatrix} C \\ CA \\ CA^2 \\ \\ CA^{r-1} \end{pmatrix} = \begin{pmatrix} \omega_{11} & \omega_{12} & .. & \omega_{1r} \\ \omega_{21} & \omega_{22} & .. & \omega_{2r} \\ .. & .. & .. & .. \\ \omega_{r1} & \omega_{r2} & .. & \omega_{rr} \end{pmatrix} \begin{pmatrix} k_g & -k_h \\ -\frac{k_g}{\tau_g} & \frac{k_h}{\tau_h} \\ \frac{k_g}{\tau_g{}^2} & -\frac{k_h}{\tau_h{}^2} \\ \\ \frac{(-1)^{r-1} k_g}{\tau_g{}^r} & \frac{(-1)^r k_h}{\tau_h{}^r} \end{pmatrix} = \begin{pmatrix} 0 & 0 \\ 0 & 0 \\ 0 & 0 \\ .. & \\ 0 & 0 \end{pmatrix}$$

*i.e,* $r^2$ *unknowns for* $2r$ *equations* (34)

*If there are solutions other than the null matrix for* $\Omega$ *then:*

$$P = \Omega\left[Y(k,r) - \mathbf{B}(r)U(k,r)\right] = \Omega\varepsilon(r)$$

*provides redundancy relationships between the measured inputs and outputs since the state has been eliminated from the generated equations. Failure to verify these redundancy relationships by the measurements acquired online will therefore be symptomatic of the occurrence of an anomaly that can be linked to, for example, a risk of injury or exhaustion.*

## 7. Modelling uncertainty in the fitness-fatigue model

In order to propose a complete panel of the contribution of the automatic tools to the sports sciences and in particular, the analysis of the relation load/performance, we insisted on reserving a section to the Kalman filter. This section about uncertainty in the Banister's model merely takes up the articles (Swinton *et al*., 2021). (Kolossa *et al,* 2017) also describes how the Kalman filter can be combined with a state-space representation of a fitness-fatigue model to obtain better estimates of fitness and fatigue with the incoming data. However, we give in appendix 2 the details of the calculations, which allow understanding the steps of the construction of the Kalman filter stated in these articles. The standard fitness-fatigue model does not attempt to model the uncertainty in the processes governing estimates of fitness, fatigue and performance. There are some limited examples where estimates are updated by applying constraints that link, for example, successive sets of parameter estimates via a least squares algorithm (Busso *et al*., 2002). However, fitness-fatigue models can be expressed as linear models in state space incorporating uncertainty. The recursive form of the standard fitness-fatigue model can be respecified as a linear model in the state representation by incorporating additive noise on the state to model uncertainty:

$$X_{n+1} = AX_n + B\omega_n + v_n \qquad (35)$$

where $v_n$ is the noise on the state characterized by a covariance matrix $2\times 2$ which we will denote by $Q = \begin{pmatrix} \sigma_g^2 & \sigma_{g,h} \\ \sigma_{g,h} & \sigma_h^2 \end{pmatrix}$ and where the state of the system is not directly observed, but accessible by means of the indirect measurement of the performance:

$$P_n = CX_n + \eta_n \qquad (36)$$

State noise $v_n$ describes random changes in state that escape the deterministic component involving training and past fitness or fatigue. (Kolossa *et al*., 2017) noted that these random changes can be considered as unmodeled athlete effort and additional characteristics that are not accounted for; $\eta_n$ is the observed performance noise described by the measurement errors whose variance is noted $\xi^2$.

Expressing the fitness-fatigue model in the state representation has the advantage of being able to model uncertainty in the state variables and measured performance by making explicit the idea that all models are necessarily bounded and contain uncertainty. The ability of the Kalman filter to operate on quantities with statistical noise over time and to update itself iteratively can provide practitioners with an effective resource for optimizing training prior to major events. The purpose of the Kalman filter is to generate an "a posteriori" state estimate $\widehat{X}_n$ based on the "a priori" estimate $Z_n$ of deterministic behavior and observed performance $P_n$ (related to the input charge $\omega_{n-1}$). The extent to which the a priori estimate is updated depends on the relative importance of the state uncertainty to the measurement uncertainty (the ratio of $Q$ to $\xi^2$ ). When the state uncertainty is large relative to the measurement uncertainty, the "Kalman gain" will be high and the filter will give more weight to the incoming performance data and relatively large corrections can be made to the a posteriori estimate. On the other hand, when the measurement uncertainty is large relative to the state uncertainty, little weight will be given to the incoming data and the correction to the initial a priori estimate will be minimal.

More formally, the steps of the Kalman filter are expressed in the following recurrent procedure:

1/ Calculation of the a priori estimate:

$$Z_n = A\widehat{X}_{n-1} + B\omega_{n-1} \tag{37}$$

where $\widehat{X}_{n-1}$ is the previous a posteriori state estimate or, in the case of $n$ = 1, $\widehat{X}_0$ is the initial state which can be set to $g(0) = h(0) = 0$ or still estimated.

2/ Calculation of the Kalman gain (here a matrix $2 \times 1$ ) defined by:

$$K_n = M_n C^T \left(\xi^2 + CM_n C^T\right)^{-1} \tag{38}$$

where $M_n$ is the covariance matrix of $\widehat{X}_n$ which is iteratively updated. The Kalman gain scales the transformed matrix $M_n C^T$ by multiplying it by the scalar $\left(\xi^2 + CM_n C^T\right)^{-1}$ , which describes the total variance of the state, plus the variance of the measurement.

3/ Calculation of the a posteriori state estimate:

$$\widehat{X}_n = Z_n + K_n\left(P_n - CZ_n\right) \tag{39}$$

The third stage of the Kalman filter is the correction stage where the a priori estimate $Z_n$ is updated after observing the performance $P_n$ . In our study, the Kalman gain $K_n$ is expressed as a 2 × 1 matrix, so that the fitness and fatigue correction can be distinct and influenced by the state of each component and the specified error covariance. The Kalman gain for each component is multiplied by a scalar that is equal to the difference between the observed performance and the initial estimated performance (*i.e*. the difference between the two). Thus, the greater the difference between the initial estimated performance and the observed performance, the greater the correction.

4/ Update of the covariance matrix of the a posteriori errors of the estimation $M_n$

$$\begin{aligned} M_{n+1} &= Q + AM_n A^T - AM_n C^T\left(\xi^2 + CM_n C^T\right)^{-1} CM_n A^T \\ &= Q + AM_n A^T - AK_n CM_n A^T \\ &= Q + A\left(I - K_n C\right)M_n A^T \end{aligned} \tag{40}$$

After the initialization, the update of $M_n$ governs the evolution of the Kalman gain over time and the intensity of the filtering effect of the model.

The four stages of the Kalman filter can then be iterated, with the estimated state and performance being updated based on the new training data and corrections applied when performance is measured. Appendix 2 provides details of the calculations that establish these equations.

# 8. Conclusion

If the FFM model was designed to best account for the relationship between training load and sport performance, the interpretability of the model since 1975 has been a more important criterion than the accuracy of the simulations. This paper is not discussing the relevance of the model, but how to use it.

The idea behind this model is to be able to determine the training that will maximize performance at a pre-determined date, minimize athlete fatigue, or best distribute the load over time. However, the sole use in simulation makes these issues inextricable combinatorial problems. Even though the genesis of the FFM model lies in the mathematical formalism of the state representation, sports scientists seem to have overlooked the functionalities offered by the state representation framework.

The previous questions should thus be related to the control of linear systems. There is no longer a combinatorial problem associated with the search for the most suitable training for the objectives of the coach or physical trainer, but simple algebraic resolutions. Similarly, this algebraic structure offers tools for eliminating measurement noise, estimating fatigue and fitness even in the case of disturbances or unknown initial values, and diagnosing the risks of injury or overtraining. These are all functional opportunities that we wanted to highlight before Banister's model in 1975 and all the models that have been developed from it until today were forgotten because sports science researchers preferred the tools of machine learning... By abandoning some of the physiological interpretation of the initial model, richer models could be constructed in the state representation formalism, albeit at the expense of the physiological interpretation, but more precise and nevertheless mathematically interpretable. Thus, the state representation could open up new horizons in terms of controllability and diagnosability for coaches and physical trainers. This study tries to give a new scope to this historical model. Furthermore, all the calculations that may seem complicated can be automated with the Matlab simulation environment without having much expertise in control theory. This article does not aim to solve the question of optimal training in practice, for that it would need to be validated by specialists in physiology and sports science, it just proposes to take a fresh look at training issues.

Machine learning will certainly outperform Banister's model in its original form, but the features outlined in this study will be lost as will the interpretability of the model. We have therefore devoted a first part of this study to show how easy it was, while keeping the mathematical formalism of the state representation (and consequently the related functionalities), to make the model more complex without any particular effort and while keeping not the physiological but the mathematical interpretability. The conclusion of this study could be: before throwing away the Banister model, has it been ensured that all its richness has been exploited?

************

# Appendix 1: Lyapunov's theorem and LQ optimal control[3]

Stability in the Lyapunov sense: The steady state $X_e = 0$ is globally asymptotically stable if there is a continuously differentiable function $V(X)$ such that: 1. $V(0)=0$ , 2. $V(X)>0, \forall X \neq 0$ , 3. $\dot{V}(X)<0, \forall X \neq 0$ and 4. $\dot{V} \to -\infty$ when $\|X\| \to \infty$.

The system $\dot{X} = \frac{dX(t)}{dt} = AX(t)$ (1) is globally asymptotically stable if all eigenvalues of $A$ are strictly negative, i.e., $\forall i = 1,\dots,n, \operatorname{Re}(\lambda_i(A)) < 0$ .

Lyapunov's theorem on the stability of linear systems states that the system defined by equation (1) is asymptotically stable if and only if for any symmetric positive definite matrix $Q$ , there exists a positive definite matrix $P$ satisfying Lyapunov's equation: $A^T P + PA + Q = 0$ (2)

[3] This appendix is a synthesis inspired by the reference https://pagespro.isae-supaero.fr/IMG/pdf/cours_note2.pdf

<u>Sufficient condition</u>: let the Lyapunov function be a candidate $V(X) = X^T PX$ then:

$$\dot{V}(X) = X^T P\dot{X} + \dot{X}^T PX = X^T PAX + X^T A^T PX = X^T (PA + A^T P)X .$$

Let $Q$ be a positive definite matrix, if $P$ is a positive solution of (2) then:

$V(X) > 0, \forall X \neq 0$ and $\dot{V}(X) = -X^T QX \Rightarrow \dot{V}(X) < 0, \forall X \neq 0$ .

Therefore, according to the general theorem, the system is asymptotically stable.

<u>Necessary condition</u>: for any given pair $(A, Q)$ , the equation (2) with unknown $P$ may not admit a unique solution, but if $A$ is stable then the Lyapunov equation admits a unique solution: $P = \int_0^\infty e^{A^T t} Q e^{At} dt$ since:

$$A^T P + PA = \int_0^\infty A^T e^{A^T t} Q e^{At} dt + \int_0^\infty e^{A^T t} Q e^{At} A dt = \int_0^\infty \frac{d}{dt}\left(e^{A^T t} Q e^{At}\right) dt = \left[e^{A^T t} Q e^{At}\right]_0^\infty = 0 - Q = -Q$$

because $e^{At} \to 0$ when $t \to \infty$ if $A$ is stable.

In what follows on the Linear-Quadratic (LQ) control, we use a variant of the stability condition: let $A$ be a stable matrix then $P$ solution of (2) is positive definite if $Q$ is positive definite.

<u>The LQ command</u>

Let the linear system be $\begin{cases} \dot{X}(t) = AX(t) + BU(t) \\ Y(t) = CX(t) \end{cases}$

The state feedback control that stabilizes the system and minimizes the LQ criterion:

$J = \int_0^{+\infty} [Y^T(t)QY(t) + U^T(t)RU(t)]dt = \int_0^{+\infty} [X^T(t)Q_X X(t) + U^T(t)RU(t)]dt$ with $R > 0, Q \geq 0$ and $R > 0,\ Q \geq 0$ and $Q_X = C^T QC$, is written:

$U(t) = -K_c X(t)$ with $K_c = R^{-1} B^T P_c$ and $P_c$ a positive solution to the Ricatti equation:

$$P_c A + A^T P_c - P_c BR^{-1} B^T P_c + Q_x = 0 .$$

Then we have: $J_{\min} = X^T(0) Q_X X(0)$.

<u>Proof</u>:

The dynamics of the closed loop system on the control law $U(t) = -KX(t)$ obeys the equation: $\dot{X}(t) = (A - BK)X(t)$.

The autonomous answer of $X(t)$ is then written: $X(t) = e^{A_f t} X_0$ with $A_f = A - BK$ and $X(0) = X_0$

The criterion $J$ becomes:

$$J = \int_0^{+\infty} [X^T(t)Q_X X(t) + U^T(t)RU(t)]dt = X_0^T \left( \int_0^{+\infty} [e^{A_f^T t}\left(Q_X + K^T RK\right)e^{A_f t}]dt \right) X_0 = X_0^T P X_0$$

with $P = \int_0^{+\infty} [e^{A_f^T t}\left(Q_X + K^T RK\right)e^{A_f t}]dt$ .

The constraint $A_f$ stable means that $P$ verifies the Lyapunov equation:

$$A_f^T P + PA_f + Q_x + K^T RK = 0$$

Moreover $P \geq 0$ as $J = X_0^T P X_0$ and $J \geq 0$ $\forall X_0$ (quadratic criterion).

Let $K_c$ be the optimal value of $K$ which minimizes $J$ and $P_c$ the corresponding solution of the Lyapunov equation:

$$\left(A - BK_c\right)^T P_c + P_c\left(A - BK_c\right) + Q_x + K_c^T RK_c = 0 \qquad (3)$$

Consider a variation $\Delta K$ around $K_c$; let $K = K_c + \Delta K$, then it results in a variation $\Delta P$ around $P_c$, and $P = P_c + \Delta P$ which verifies:

$$\left(A - B\left(K_c + \Delta K\right)\right)^T\left(P_c + \Delta P\right) + \left(P_c + \Delta P\right)\left(A - B\left(K_c + \Delta K\right)\right) + Q_x + \left(K_c + \Delta K\right)^T R\left(K_c + \Delta K\right) = 0 \qquad (4)$$

$K_c$ is the optimal value in the sense of the criterion $J$ if and only if the criterion increases for any variation $\Delta K$ around $K_c$, *i.e.*:

$\Delta P > 0$ $\forall \Delta K$ / $A - B\left(K_c + \Delta K\right)$ stable.

(if $\Delta K$ is such that $A - B\left(K_c + \Delta K\right)$ is unstable, then the criterion becomes infinite).

Subtracting (3) from (4) gives:

$$\left(A - B\left(K_c + \Delta K\right)\right)^T\left(P_c + \Delta P\right) + \left(P_c + \Delta P\right)\left(A - B\left(K_c + \Delta K\right)\right) + Q_x + \left(K_c + \Delta K\right)^T R\left(K_c + \Delta K\right)$$

$$-\left(A - BK_c\right)^T P_c + P_c\left(A - BK_c\right) + Q_x + K_c^T RK_c = 0$$

Or:

$$\left(A - BK\right)^T \Delta P + \Delta P\left(A - BK\right) + \Delta K^T\left(RK_c - B^T P_c\right) + \left(RK_c - B^T P_c\right)^T \Delta K + \Delta K^T R\Delta K = 0$$

This is a Lyapunov equation. $\left(A - BK\right)$ being stable, $\Delta P$ is positive if and only if:

$$\Delta K^T\left(RK_c - B^T P_c\right) + \left(RK_c - B^T P_c\right)^T \Delta K + \Delta K^T R\Delta K > 0 \quad \forall \Delta K$$

Now $\Delta K^T R\Delta K > 0$ $\forall \Delta K$ because $R > 0$ and it is therefore necessary that:

$RK_c - B^T P_c = 0$ or $K_c = R^{-1}B^T P_c$ .

If we transfer this equation to (3), we obtain the Ricatti equation:

$$\left(A-BR^{-1}B^TP_c\right)^TP_c+P_c\left(A-BR^{-1}B^TP_c\right)+Q_x+\left(R^{-1}B^TP_c\right)^TRR^{-1}B^TP_c=0$$
$$\Rightarrow P_cA+A^TP_c-P_c\,BR^{-1}B^TP_c+Q_x=0$$

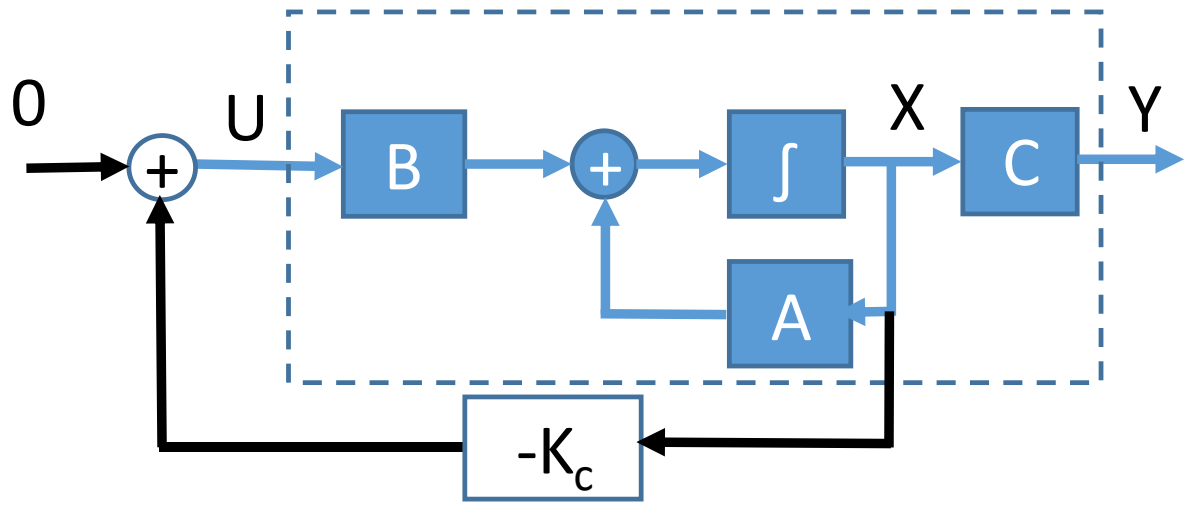

*Schematic diagram of LQ control*

************

## Appendix 2: Kalman Filter Recurrence Equations[4]

Notations and preliminary results.

Let us consider two random vectors $X\in\mathbb{R}^n$ and $Y\in\mathbb{R}^m$. The vector $Y$ corresponds to the vector of measurements and $X$ to the state to be estimated.

An estimator is a function $\phi(Y)$ that is supposed to give an estimate of $X$ knowing the measure $Y$.

The analytical expression for such an estimator is generally not easy to obtain and we generally prefer to limit ourselves to linear estimators. A linear estimator is a linear $\mathbb{R}^m\rightarrow\mathbb{R}^n$ function of the form:

$$\widehat{X}=KY+b$$

where $K\in\mathbb{R}^{n\times m}$ and $b\in\mathbb{R}^n$. The idea is to find a *good* $K$ and a *good* $b$ from the sole knowledge of the first order moments $\overline{X},\overline{Y}$ and the second order moments $\Gamma_x,\Gamma_y,\Gamma_{xy}$.

The estimation error is:

$$\varepsilon=\widehat{X}-X$$

The estimator is said to be unbiased if $E(\varepsilon)=0$. It is said to be orthogonal if $E\left(\varepsilon\tilde{Y}^T\right)=0$ [5] with $\tilde{Y}=Y-E(Y)=Y-\overline{Y}$.

Theorem: Let be two random vectors $X$ and $Y$. There is a unique orthogonal unbiased estimator and it is given by:

$$\widehat{X}=\overline{X}+K\left(Y-\overline{Y}\right)\text{ with }K=\Gamma_{XY}\Gamma_Y^{-1}$$

$K=\Gamma_{XY}\Gamma_Y^{-1}$ is called the Kalman gain.

---

[4] This appendix is a synthesis inspired by the reference https://www.ensta-bretagne.fr/jaulin/poly_kalman.pdf

[5] The name comes from a scalar product on the space of random variables: $\langle a,b\rangle=E\left(\left(a-\overline{a}\right)\left(b-\overline{b}\right)\right)$.

Indeed,

$$E(\varepsilon)=E(\widehat{X}-X)=E(KY+b-X)=KE(Y)+b-E(X)=K\overline{Y}+b-\overline{X}$$

The estimator is unbiased if $E(\varepsilon)=0$ or: $b=\overline{X}-K\overline{Y}$

In this case:

$$\varepsilon=\widehat{X}-X=KY+b-X=KY+\overline{X}-K\overline{Y}-X=K\tilde{Y}-\tilde{X} \text{ with } \tilde{Z}=Z-\overline{Z}$$

The estimator is orthogonal if:

$$E(\varepsilon\tilde{Y}^T)=0\Leftrightarrow E\left(\left(K\tilde{Y}-\tilde{X}\right)\tilde{Y}^T\right)=0\Leftrightarrow E\left(K\tilde{Y}\tilde{Y}^T-\tilde{X}\tilde{Y}^T\right)=0\Leftrightarrow K\Gamma_Y-\Gamma_{XY}=0\Leftrightarrow K=\Gamma_{XY}\Gamma_Y^{-1}$$

Thus, we have:

$$\widehat{X}=KY+b=KY+\overline{X}-K\overline{Y}=\overline{X}+K(Y-\overline{Y})=\overline{X}+\Gamma_{XY}\Gamma_Y^{-1}(Y-\overline{Y})$$

The covariance matrix of the error of the linear orthogonal unbiased estimator is:

$$\Gamma_\varepsilon=\Gamma_X-K\Gamma_{YX}$$

Indeed:

$$\begin{aligned}\Gamma_\varepsilon&=E(\varepsilon\varepsilon^T)=E\left(\left(K\tilde{Y}-\tilde{X}\right)\left(K\tilde{Y}-\tilde{X}\right)^T\right)\\&=E\left(\left(K\tilde{Y}-\tilde{X}\right)\left(\tilde{Y}^TK^T-\tilde{X}^T\right)\right)\\&=E\left(K\tilde{Y}\tilde{Y}^TK^T-\tilde{X}\tilde{Y}^TK^T-K\tilde{Y}\tilde{X}^T+XX^T\right)\end{aligned}$$

Or: $\Gamma_\varepsilon=K\Gamma_YK^T-\Gamma_{XY}K^T-K\Gamma_{YX}+\Gamma_X$

In the case where the estimator is orthogonal, $K=\Gamma_{XY}\Gamma_Y^{-1}$ and we obtain: $\Gamma_\varepsilon=\Gamma_X-K\Gamma_{YX}$

There is no unbiased linear estimator that gives a smaller covariance matrix on the error $\Gamma_\varepsilon$ than that given by the orthogonal estimator.

Application to linear filtering

Let us assume that two quantities $X$ and $Y$ are related by the relationship:

$$Y=CX+\beta$$

where $\beta$ is a centered random vector, uncorrelated with X. The covariance matrices of $X$ and $\beta$ are denoted $\Gamma_X$ and $\Gamma_\beta$. The best-unbiased linear estimator for $X$ is sought.

We have:

$$\overline{Y}=C\overline{X}+\overline{\beta}=C\overline{X}$$

$$\Gamma_Y = E\left(\left(Y-\bar{Y}\right)\left(Y-\bar{Y}\right)^T\right) = E\left(\left(CX+\beta-C\bar{X}\right)\left(CX+\beta-C\bar{X}\right)^T\right)$$

$$= E\left(\left(C\tilde{X}+\beta\right)\left(C\tilde{X}+\beta\right)^T\right) = CE\left(\tilde{X}\tilde{X}^T\right)C^T + CE\underbrace{\left(\tilde{X}\beta^T\right)}_{=0} + E\underbrace{\left(\beta\tilde{X}^T\right)}_{=0}C^T + E\left(\beta\beta^T\right)$$

or: $\Gamma_Y = C\Gamma_X C^T + \Gamma_\beta$

$$\Gamma_{XY} = E\left(\tilde{X}\tilde{Y}^T\right) = E\left(\tilde{X}\left(C\tilde{X}+\tilde{\beta}\right)^T\right) = E\left(\tilde{X}\tilde{X}^T C^T + \tilde{X}\tilde{\beta}^T\right) = E\left(\tilde{X}\tilde{X}^T\right)C^T + E\underbrace{\left(\tilde{X}\tilde{\beta}^T\right)}_{=0}$$

or: $\Gamma_{XY} = \Gamma_X C^T$ .

The best-unbiased estimator for $X$ can then be obtained from $\Gamma_X, \Gamma_\beta, C, \bar{X}$ using the following formulas:

1) Estimated: $\hat{X} = \bar{X} + K\tilde{Y}$
2) Covariance of the error: $\Gamma_\varepsilon = \Gamma_X - K\Gamma_{YX} = \Gamma_X - KC\Gamma_X$
3) Innovation factor: $\tilde{Y} = Y - C\bar{X}$
4) Covariance of innovation: $\Gamma_Y = C\Gamma_X C^T + \Gamma_\beta$
5) Kalman gain: $K = \Gamma_X C^T \Gamma_Y^{-1}$

Consider the following dynamic system:

$$\begin{cases} X_{k+1} = AX_k + U_k + \alpha_k \\ Y_k = CX_k + \beta_k \end{cases}$$

where $\alpha_k$ and $\beta_k$ are mutually independent and white Gaussian random signals[6] in time. The Kalman filter alternates two phases: correction and prediction. To understand the mechanism of such a filter, let us place ourselves at the time $k$ and suppose that we have already processed the measurements $Y_0, Y_1, ..., Y_{k-1}$. At this stage, the state vector is a random vector $X_{k/k-1}$ (at the time $k$ and the measurements have been processed up to $k-1$). This random vector is denoted $\hat{X}_{k/k-1}$ and the associated covariance matrix is $\Gamma_{k/k-1}$ .

<u>Kalman filter recurrence equations</u>

| | |
|---|---|
| Predicted estimate: | $\hat{X}_{k+1/k} = A\hat{X}_{k/k} + U_k$ |
| Predicted covariance: | $\Gamma_{k+1/k} = A\Gamma_{k/k}A^T + \Gamma_{\alpha_k}$ |
| Corrected estimate: | $\hat{X}_{k/k} = \hat{X}_{k/k-1} + K_k\tilde{Y}_k$ |
| Corrected covariance | $\Gamma_{k+1/k} = \left(I - K_k C\right)\Gamma_{k/k-1}$ |
| Innovation | $\tilde{Y}_k = Y_k - C\hat{X}_{k/k-1}$ |
| Innovation covariance | $S_k = C\Gamma_{k/k-1}C^T + \Gamma_{\beta_k}$ |
| Kalman gain: | $K_k = \Gamma_{k/k-1}C^T S_k^{-1}$ |

[6] *White* means that the vectors $\alpha_{k_1}$ and $\alpha_{k_2}$ are independent of each other.